\documentclass[aps,prb,10pt,a4paper,twocolumn]{revtex4-2}
\usepackage{graphicx,overpic}
\usepackage{amssymb,amsmath,mathtools,textcomp}
\usepackage{bbm,enumerate}
\usepackage{subcaption}
\usepackage[newcommands]{ragged2e}
\usepackage{tabularx,colortbl}
\usepackage{array}
\usepackage{multirow}
\usepackage{comment}
\usepackage{ bbold }

\usepackage{soul}
\usepackage{appendix}

\newcommand{\conb}{CoNb$_2$O$_6$}

\newcommand{\Qinc}{\vec{Q}_{\rm inc}}
\newcommand{\qinc}{q_{\rm inc}}

\newcommand{\llangle}{\langle\!\langle}
\newcommand{\rrangle}{\rangle\!\rangle}

\graphicspath{{./}{./figs/}}

\begin{document}

\title{
Field-driven phases in a three-dimensional twisted Kitaev model for CoNb$_2$O$_6$: \\
Interplay of frustration and spin-orbit coupling
}
\author{Tom Drechsler}
\author{Matthias Vojta}
\affiliation{Institut f\"ur Theoretische Physik and W\"urzburg-Dresden Cluster of Excellence ctd.qmat, Technische Universit\"at Dresden, 01062 Dresden, Germany}

\begin{abstract}
The Ising chain in a transverse field stands out as a paradigmatic example for a quantum phase transition. {\conb} has been discussed as a material realization of this physics, but it was later realized that its magnetic exchange couplings are more complicated, taking the form of twisted Kitaev chains. Here we study a three-dimensional model for CoNb$_2$O$_6$, taking into account both Kitaev physics and frustrated interchain coupling, in applied magnetic field. Using semiclassical techniques at zero temperature, we map out the sequence of field-driven phases for arbitrary field direction; these include phases with commensurate and incommensurate interchain order. As a result of spin-orbit coupling, the phase diagram is extremely sensitive to small changes in the field angle. We compute static observables as well as magnetic excitation spectra in the various phases and connect our results to existing experimental data.
\end{abstract}

\date{\today}

\maketitle


\section{Introduction}

The transverse-field Ising model is a textbook example for a quantum phase transition in magnetic insulators: Easy-axis magnetism can be destroyed by a magnetic field applied perpendicular to the easy axis \cite{book_sachdev}. A number of materials realizations have been discussed, among them LiHoF$_4$ \cite{bitko96}, CoNb$_2$O$_6$ \cite{maartense77,coldea10}, and (Ba/Sr)Co$_2$V$_2$O$_8$ \cite{he05,wang18,cui19}. Interestingly, each of these materials comes with its own intricacies:
For instance, LiHoF$_4$ is a ferromagnet dominated by three-dimensional dipolar interactions between local moments of non-Kramers nature, and moreover features a large hyperfine coupling to nuclear spins. As a result, the critical behavior is mean-field-like both at finite and zero temperature, the phase diagram is reshaped by hyperfine effects, and the low-energy physics near quantum criticality is dominated by electro-nuclear modes \cite{bitko96,ronnow05,wendl22,eisenlohr21}.

The present paper is devoted to CoNb$_2$O$_6$ where, in contrast, the ferromagnetic interactions are predominantly one-dimensional, and the coupling between these chains is antiferromagnetic and frustrated. Therefore, signatures of one-dimensional Ising criticality have been found at elevated temperatures and energies \cite{coldea10,kinross14}, whereas three-dimensional antiferromagnetic phases prevail at low $T$ \cite{scharf79,heid95,heid97}. In addition, it has been argued that the combination of low crystal symmetry and spin-orbit coupling renders the interactions more complicated than that of a simple Ising model. Instead, the picture of bond-dependent Ising interactions has been put forward \cite{morris21}, somewhat similar to those of the honeycomb-lattice Kitaev model \cite{kitaev06}; the corresponding model has been dubbed twisted Kitaev chain. However, this is still a rather simplified model, and more elaborate models have been derived from either ab-initio calculations or fits to experimental data. While much attention has been paid to the intrachain interactions, it is the frustrated interchain coupling which is responsible for the plethora of low-temperature ordered phases \cite{lee10}, and no comprehensive modeling is currently available.

The purpose of this paper is to close this gap. We study three-dimensional spin models for CoNb$_2$O$_6$ and determine the ground-state phases as function of applied magnetic field for arbitrary field direction. Our analysis employs semiclassical techniques which capture the properties of the low-$T$ ordered phases in a semi-quantitative fashion. We utilize different sets of model parameters from the literature and discuss the physics upon variation of the interchain couplings. We find multiple antiferromagnetic phases with both commensurate and incommensurate order, with a surprisingly complex evolution as function of the field strength and direction. We provide concrete predictions for the evolution of magnetic order as well as both the longitudinal and transverse magnetization components upon field rotation. We also determine the low-energy spin-wave excitations across the phase diagram, and we discuss freezing effects seen near the quantum phase transitions of CoNb$_2$O$_6$.

The remainder of the paper is organized as follows:
In Sec.~\ref{sec:model} we discuss the microscopic model and its symmetries. Sec.~\ref{sec:swt} summarizes the semiclassical approach employed to access the low-$T$ ordered phases. Its results are contained in the following sections: Sec.~\ref{sec:phases} describes the sequence of ordered phases and the associated phase transitions induced by a magnetic field, while Sec.~\ref{sec:mag} highlights the features of the uniform magnetization, both along and perpendicular to the applied field. Sec.~\ref{sec:exc} presents the excitation spectra obtained within linear spin-wave theory. Finally, Sec.~\ref{sec:comp} provides a discussion of our results vis-a-vis experimental data, also including physics beyond the semiclassical equilibrium model employed here, such as the experimentally observed freezing.
A brief summary of our results, together with suggestions for future work, closes the paper.
\section{Exchange model for $\mathrm{Co}\mathrm{Nb}_2\mathrm{O}_6$}
\label{sec:model}
As the low-temperature behavior can be expected to depend sensitively on details of the microscopic model, we describe its ingredients step by step in the following.

\begin{figure}[tb]
\includegraphics[width=\columnwidth]{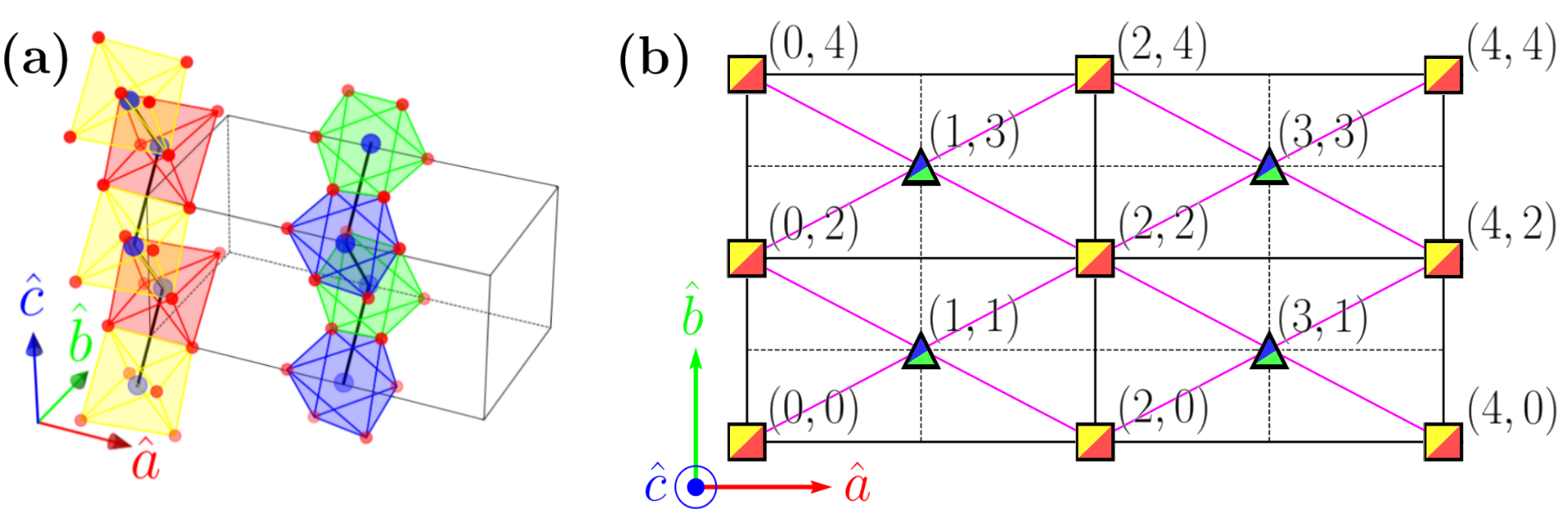}
\caption{
Crystal structure of \conb. \textbf{(a):} The $\mathrm{Co}^{2+}$ ions (blue) are surrounded by distorted edge-sharing octahedra of $\mathrm{O}^{2-}$ ions (red) and form chains running along the $c$ axis. The $\mathrm{Nb}^{2+}$ ions and additional $\mathrm{O}^{2-}$ ions between the chains are not shown. The crystallographic unit cell contains two inequivalent zigzag chains running in the $c$ direction. Both chains are related by an $n$ glide. \textbf{(b):} In the $ab$ plane, the chains ($\square$, $\Delta$) form acute isosceles triangles (magenta) and are labeled by a double index $(i,i')$, see text.
}
\label{fig:crystal_structure_CoNb2O6}
\end{figure}

\conb{} crystallizes in the orthorhombic columbite structure \cite{ringler22}. The $\mathrm{Co}^{2+}$ ions, surrounded by six $\mathrm{O}^{2-}$ ions, form zigzag chains of slightly distorted edge-sharing octahedra running in the $c$ direction. The crystal structure hosts two types of inequivalent chains which are related by a glide symmetry, Fig.~\ref{fig:crystal_structure_CoNb2O6}(a).

The crystalline electric field (CEF) leads to a splitting of the 28-fold degenerate Co ${}^4F$ multiplet into doublets \cite{ringler22}. The ground state is a Kramers doublet, separated from the next CEF levels by a splitting of about 30\,meV. This allows for a description in terms of an effective spin-$1/2$ model, which has been developed and discussed for both isolated and coupled chains in numerous papers \cite{ringler22,heid95,morris21,lee10,robinson14,cabrera14,fava20,thota21,woodland23I,woodland23II,gallegos24,churchill24,birnkammer24,konieczna25}. A general bilinear Hamiltonian with nearest-neighbor interaction between pseudospins $\hat{\vec{S}}_j^{(i,i')}$ reads
\begin{align}
\hat{H}_{\mathrm{NN}|\mathrm{chains}}
=
\sum_{ii'}\sum_{j} \hat{\vec{S}}_j^{(i,i')}{}^{\mathrm{T}} \cdot \mathbf{M}^{(i,i')}_{j,j+1} \cdot \hat{\vec{S}}_{j+1}^{(i,i')}
\label{eq:nearest_neighbor_chain}
\end{align}
where $j$ runs along $c$ and refers to the sites along the chains. The latter are enumerated by a set of $ab$ coordinates $(i,i')$ chosen such that $i$ and $i'$ are either both even or both odd, as illustrated in Fig.~\ref{fig:crystal_structure_CoNb2O6}(b).
The matrix $\mathbf{M}^{(i,i')}_{j,j+1}$, here written in a global reference frame spanned by the crystallographic axes $(\hat{a},\hat{b},\hat{c})$, contains all symmetric nearest-neighbor couplings in chain $(i,i')$; antisymmetric Dzyaloshinskii–Moriya interactions are forbidden due to the presence of an inversion center between the $\mathrm{Co}$--$\mathrm{Co}$ bonds \cite{konieczna25}.

To our knowledge it is not clear whether the nuclear spins of Co are important to the magnetism of {\conb} at experimentally relevant temperatures. Ref.~\onlinecite{liang15} estimated the hyperfine contribution to the specific heat and concluded it to be negligible down to 300\,mK. Our modeling will therefore not include nuclear spins.

\begin{figure}[!t]
\centering
\includegraphics[width=0.28\textwidth]{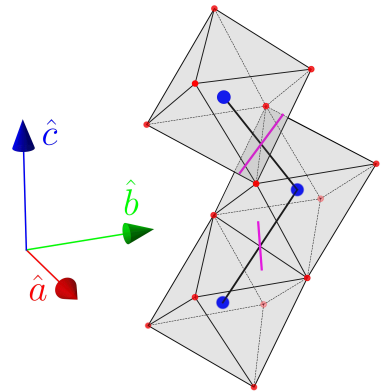}
\caption{
Single twisted Kitaev chain with two local Ising axes $\hat{n}_1^{(i,i')},\hat{n}_2^{(i,i')}$ (magenta) alternating along the chain. The axis $\hat{n}^{(i,i')}$ bisecting $\hat{n}_1^{(i,i')},\hat{n}_2^{(i,i')}$ is the axis of spontaneous magnetization of chain $(i,i')$ at $B=0$ \cite{morris21}, see text.
}
\label{fig:Single_Chain_Ising_Axes}
\end{figure}

\subsection{Twisted Kitaev chains}
\label{sec:model:Twisted_Kitaev}

Early neutron diffraction studies have shown that \conb{} has Ising-like properties, but appears to deviate from ideal Ising behavior \cite{maartense77, scharf79, heid95}. A model resolving this dichotomy has been introduced in Ref.~\onlinecite{morris21}, dubbed the twisted Kitaev chain: each Co--Co bond has a local Ising axis whose direction is dictated by the crystalline environment. These local Ising axes are not independent:
The space group $Pbcn$ contains two glide symmetries, (i) a $c$ glide which transforms the local Ising axis of one bond in chain $(i,i')$ to the Ising axis of its neighbor bond in the same chain and (ii) an $n$ glide which transforms the two chains $(i,i')\leftrightarrow (i+1,i'+1)$ inside the crystallographic unit cell into each other. As a result, each chain $(i,i')$ has two distinct (but symmetry-related) Ising axes $\hat{n}_{1}^{(i,i')},\hat{n}_2^{(i,i')}$ alternating along the chain, and the Ising axes in the chain $(i+1,i'+1)$ are symmetry-related to those in chain $(i,i')$.
The alternating Ising axes for each chain, Fig.~\ref{fig:Single_Chain_Ising_Axes}, can thus be parameterized by only two angles as follows \cite{morris21},
\begin{align}
\hat{n}_{1}^{(i,i')} &= \cos\theta\sin\phi\,\hat{a} + \sin\theta\,\hat{b} + (-1)^{i}\cos\theta\cos\phi\,\hat{c},
\notag \\
\hat{n}_{2}^{(i,i')} &= \cos\theta\sin\phi\,\hat{a} - \sin\theta\,\hat{b} + (-1)^{i}\cos\theta\cos\phi\,\hat{c},
\label{eq:local_Ising_axes}
\end{align}
which defines $\theta$ and $\phi$. The twisted Kitaev-chain Hamiltonian then reads
\begin{align}
\hat{H}_{\mathrm{NN}|\mathrm{chains}}
=
-K
&\sum_{ii'}\sum_{j} \left[(\hat{\vec{S}}_{2j}^{(i,i')}\cdot\hat{n}_1^{(i,i')})(\hat{\vec{S}}_{2j+1}^{(i,i')}\cdot\hat{n}_1^{(i,i')})
\right.
\nonumber
\\
&\left.
+
(\hat{\vec{S}}_{2j+1}^{(i,i')}\cdot\hat{n}_2^{(i,i')})(\hat{\vec{S}}_{2j+2}^{(i,i')}\cdot\hat{n}_2^{(i,i')})
\right]
\label{eq:Morris_Twisted_Kitaev}
\end{align}
where $K>0$ is the ferromagnetic Kitaev coupling. Using Eq.~\eqref{eq:local_Ising_axes}, this Hamiltonian can be brought into the general form of Eq.~\eqref{eq:nearest_neighbor_chain}, with $\mathbf{M}^{(i,i')}_{j,j+1}$ given by

\begin{align}
\mathbf{M}^{(i,i')}_{j,j+1}
=
\begin{pmatrix}
M_{aa} & (-1)^{j} M_{ab} & (-1)^{i} M_{ac} \\
(-1)^{j} M_{ba} & M_{bb} & (-1)^{i+j} M_{bc} \\
(-1)^{i} M_{ca} & (-1)^{i+j} M_{cb} & M_{cc}
\end{pmatrix}
\label{eq:Nearest_Neighbor_Matrix}
\end{align}
and
\begin{align}
M_{aa} &= -K\cos^2\theta\sin^2\phi ,\notag \\
M_{ab} &= -K\sin\theta\cos\theta\sin\phi = M_{ba} ,\notag \\
M_{ac} &= -K\cos^2\theta\sin\phi\cos\phi = M_{ca} ,\notag \\
M_{bb} &= -K\sin^2\theta ,\notag \\
M_{bc} &= -K\sin\theta\cos\theta\cos\phi = M_{cb} ,\notag \\
M_{cc} &= -K\cos^2\theta\cos^2\phi
\label{eq:Morris_Couplings}
\end{align}
are the couplings with respect to the crystallographic reference frame. The alternating prefactors $(-1)^j$, $(-1)^i$ are dictated by the crystal symmetries \cite{fava20, gallegos24}:
The $c$ glide along each chain reverses $\hat{b}\mapsto -\hat{b}$, leading to the factors $(-1)^j$ in those entries containing the index $b$ only once, while the $n$ glide between the chains reverses $\hat{c}\mapsto -\hat{c}$, producing factors $(-1)^i$ in those entries containing the index $c$ only once.

The picture of two alternating local Ising axes per chain is sufficient to qualitatively understand the behavior of the single chain in zero field \cite{morris21}: Since the Ising axes \eqref{eq:local_Ising_axes} are \textit{distinct}, the Ising interaction within each chain is frustrated (although less frustrated than in the honeycomb-lattice Kitaev model \cite{kitaev06} with its orthogonal Ising axes). As a result, each chain tends to order ferromagnetically, with the axis of spontaneous magnetization for chain $(i,i')$ given by
\begin{align}
\hat{n}^{(i,i')} \coloneq \sin\phi\,\hat{a} + (-1)^{i}\cos\phi\,\hat{c}.
\label{eq:axis_of_magnetization_single_chain}
\end{align}
This easy axis lies in the $ac$ plane and bisects the two Ising axes $\hat{n}_{1,2}^{(i,i')}$. Since the easy axes for both type of chains have a vanishing $b$ component as a result of the crystal symmetries, an external field $\vec{B}\parallel\hat{b}$ is perpendicular to the spontaneous magnetization of both the chains. This defines $\vec{B}\parallel\hat{b}$ as the transverse field direction and justifies that the transverse-field Ising model is an acceptable approximation of \conb{}. The angle ${\phi^{(i,i')}\coloneq\sphericalangle(\hat{n}^{(i,i')},\hat{c})}$ is accessible in neutron-diffraction experiments, and there is agreement among different studies that ${\phi^{(i,i')}\approx (-1)^i\, 30^{\circ}}$ \cite{maartense77,scharf79,heid95,kobayashi99,kobayashi00,kobayashi04,sarvezuk11II,churchill24}, cf.~Table~\ref{tab:Ising_axis_canting_angle}. The only free parameter left in Eq.~\eqref{eq:Morris_Couplings} is therefore the angle $\theta$. Ref.~\onlinecite{morris21} determined ${\theta\approx 17^{\circ}}$ by fitting their model to data obtained from THz spectroscopy.

\begin{table}[tb]
\begin{tabular}{|l|l|}
\hline
$|\phi^{(i,i')}|$ & Reference \\\hline
$29.6^{\circ}$ & \cite{heid95} \\\hline
$31^{\circ}$ & \cite{scharf79, kobayashi99, kobayashi00, kobayashi04, sarvezuk11II}\\\hline
$33^{\circ}$ & \cite{churchill24} \\\hline
$34^{\circ}$ & \cite{maartense77}\\\hline
\end{tabular}
\caption{
Canting angle ${\phi^{(i,i')}\coloneq\sphericalangle(\hat{n}^{(i,i')},\hat{c})}$ between the axis $\hat{n}^{(i,i')}$ of spontaneous magnetization in chain $(i,i')$ and the crystallographic $c$ axis. Throughout the literature, there is agreement that ${\phi^{(i,i')}\approx (-1)^i\, 30^{\circ}}$, with the precise value depending on the model and/or the applied fitting method.
}
\label{tab:Ising_axis_canting_angle}
\end{table}

\subsection{Bond-dependent anisotropies}
\label{sec:model:anisotropies}

The model of the twisted Kitaev chain, Eq.~\eqref{eq:Morris_Twisted_Kitaev}, exclusively involves bond-dependent Ising interactions. However, other interactions are clearly symmetry-allowed, and a general model will contain multiple energy scales, which may be parameterized, e.g., via Heisenberg, Kitaev, and $\Gamma$ interactions in analogy with Kitaev materials \cite{jackeli2009,trebst2017,winter2017b,janssen19}. The parameters of such generalized models have been extracted from fits to {\conb} experimental data \cite{robinson14, cabrera14, fava20, thota21, woodland23I, woodland23II, gallegos24, churchill24}. As discussed, e.g., in Ref.~\onlinecite{woodland23I}, nearest-neighbor models \cite{fava20, churchill24} allow one to semi-quantitatively capture most experimental features, while additional further-neighbor interactions \cite{robinson14, cabrera14, woodland23I, woodland23II, gallegos24} improve the quantitative agreement.

In addition, the interaction parameters were also extracted \cite{konieczna25} from first-principles density-functional calculations via the intermediate step of a multi-band Hubbard Hamiltonian whose numerical solution was compared to experimental data. Although the underlying microscopic Hamiltonian only accounts for nearest-neighbor interactions, the spectra from inelastic neutron scattering (INS) and THz spectroscopy matched the experiments semi-quantitatively, and the extracted couplings are in good agreement with the ones determined in Ref.~\onlinecite{woodland23II}.

In what follows, we will employ the microscopic couplings from the DFT model by Konieczna \textit{et al.} \cite{konieczna25} (referenced by the authors as "Model$^{\mathrm{DFT}}_{\mathrm{corr}}$"). The results will be compared to the ones obtained from the experimentally fitted Hamiltonian by Woodland \textit{et al.} \cite{woodland23II}. Both models include Heisenberg, Kitaev, and $\Gamma$ interactions, see Appendix \ref{app:coupling_matrices} for the corresponding conversion.
We note that dipole--dipole interactions are present as well, but will be assumed to be negligible \cite{weitzel00}.

In this subsection, we are concerned with the intrachain interactions:
Ref.~\onlinecite{konieczna25} uses a special local reference frame $(\hat{x}^{(i,i')},\hat{y}^{(i,i')},\hat{z}^{(i,i')})$ for each Co---Co chain $(i,i')$ to parametrize the couplings which makes the glide symmetries transparent. We convert their Hamiltonian to the global frame where it takes the same form as in Eqs.~\eqref{eq:nearest_neighbor_chain}, \eqref{eq:Nearest_Neighbor_Matrix}, with the matrix entries $M_{\mu\nu}$ given in Table~\ref{tab:chain_couplings}.
Ref.~\onlinecite{woodland23II} uses a third reference frame $(\hat{X}^{(i,i')},\hat{Y}^{(i,i')},\hat{Z}^{(i,i')})$, given by the principal axes of the anisotropic $g$ tensor, cf.~Sec.~\ref{sec:model:g_tensor}. They also include a second-neighbor coupling, such that their intrachain Hamiltonian reads
\begin{align}
\hat{H}_{\mathrm{chains}}
=
\hat{H}_{\mathrm{NN}|\mathrm{chains}}
+
\hat{H}_{\mathrm{NNN}|\mathrm{chains}}.
\label{eq:Woodland_Chains}
\end{align}
The first-neighbor term can be brought into the form \eqref{eq:nearest_neighbor_chain}, \eqref{eq:Nearest_Neighbor_Matrix}, with matrix elements $M_{\mu\nu}$ given in Table~\ref{tab:chain_couplings}. The second-neighbor term takes the form

\begin{align}
\hat{H}_{\mathrm{NNN}|\mathrm{chains}}
=
\sum_{i,i'}\sum_{j} \hat{\vec{S}}_j^{(i,i')}{}^{\mathrm{T}} \cdot \mathbf{N}^{(i,i')}_{j,j+2} \cdot \hat{\vec{S}}_{j+2}^{(i,i')}
\label{eq:Woodland_NNN_Chains}
\end{align}
in the global frame $(\hat{a},\hat{b},\hat{c})$, with the coupling matrix
\begin{align}
\mathbf{N}^{(i,i')}_{j,j+2}
=
\begin{pmatrix}
N_{aa} & N_{ab} & (-1)^i N_{ac} \\
N_{ab} & N_{bb} & (-1)^i N_{bc} \\
(-1)^i N_{ac} & (-1)^i N_{bc} & N_{cc}
\end{pmatrix}
\label{eq:Woodland_NNN_Matrix}
\end{align}
this time \textit{not} containing alternating prefactors $(-1)^j$, because the two sites $j$ and $(j+2)$ are connected by the product of two glide symmetries. The numerical values $N_{\mu\nu}$ are also given in Table~\ref{tab:chain_couplings}.
It is worth emphasizing that for both parameter sets the couplings remain dominant Ising-like.

\begin{table}
\begin{tabular}{|r||r|r|r|}
\hline
$\mu\nu$ & $M_{\mu\nu}$ \cite{konieczna25} & $M_{\mu\nu}$ \cite{woodland23II} & $N_{\mu\nu}$ \cite{woodland23II} \\\hline
$aa$ & $-0.8200$ & $-1.0496$ & $+0.1058$\\\hline
$ab$ & $-0.0086$ & $+0.2808$ & $0$\\\hline
$ac$ & $+0.7573$ & $+0.8270$ & $+0.0497$\\\hline
$bb$ & $-0.4235$ & $-0.6768$ & $+0.0772$\\\hline
$bc$ & $-0.4443$ & $-0.4864$ & $0$\\\hline
$cc$ & $-1.9645$ & $-2.0045$ & $+0.1632$\\\hline
\end{tabular}
\caption{
Matrix entries for the first-neighbor and second-neighbor exchange couplings, $M_{\mu\nu}$ and $N_{\mu\nu}$, respectively, in meV. The values have been taken from Refs.~\onlinecite{konieczna25, woodland23II}, converted into the global crystallographic reference frame $(\hat{a},\hat{b},\hat{c})$ and rounded to four digits of precision.
}
\label{tab:chain_couplings}
\end{table}

\subsection{Anisotropic $g$ tensor}
\label{sec:model:g_tensor}

The Zeeman coupling of the pseudospins to an external magnetic field $\vec{B}$ can be written as
\begin{align}
\hat{H}_{\mathrm{field}}
=
-\mu_{\mathrm{B}}\,2S \sum_{ii'}\sum_{j} \hat{\vec{S}}_j^{(i,i')}{}^{\mathrm{T}} \cdot \mathbf{g}^{(i,i')}_{j} \cdot \vec{B}
\label{eq:B_field_and_g_tensor}
\end{align}
where $\mathbf{g}^{(i,i')}_{j}$ is the $g$ tensor of the $\mathrm{Co}^{2+}$ ion at site $j$ in chain $(i,i')$ and $\mu_{\mathrm{B}}$ the Bohr magneton. The factor $2S$, equaling unity for spins $1/2$, is included for bookkeeping purposes in spin-wave theory only.
The local $g$ tensor, originating from the (distorted) octahedral environment of the $\mathrm{Co}^{2+}$ ions, alternates along and in between the chains, with the glide symmetries imposing the sign structure
\begin{align}
\mathbf{g}^{(i,i')}_{j}
=
\begin{pmatrix}
g_{aa} & (-1)^{j} g_{ab} & (-1)^{i} g_{ac} \\
(-1)^{j} g_{ba} & g_{bb} & (-1)^{i+j} g_{bc} \\
(-1)^{i} g_{ca} & (-1)^{i+j} g_{cb} & g_{cc}
\end{pmatrix}.
\label{eq:g_tensor}
\end{align}
The crystal symmetries dictate that one principal axis $(\hat{Y}^{(i,i')})$ is fixed to align with the $b$ axis \cite{robinson14, ringler22}. This fact, together with the symmetry ${\mathbf{g}=\mathbf{g}^{\mathrm{T}}}$, enforces ${g_{ab}=g_{ba} = g_{cb}=g_{bc}=0}$; it also implies that
$\vec{B}\parallel\hat{b}$ is a transverse-field direction for all chains despite the alternating $g$ tensor.

The tensor's entries $g_{\mu\nu}$ have been fitted to experimental data in Refs.~\onlinecite{kunimoto99, ringler22, woodland23I, woodland23II, gallegos24, konieczna25}. Throughout the literature \cite{ringler22, woodland23II, gallegos24, konieczna25} there is agreement that the principal axis $\hat{Z}^{(i,i')}$ of the largest eigenvalue in chain $(i,i')$ is close to the axis $\vec{n}^{(i,i')}$ of spontaneous magnetization.

In Table~\ref{tab:g_values} we quote the elements of the $g$ tensor from Refs.~\onlinecite{konieczna25,woodland23II}. For Ref.~\onlinecite{konieczna25} we obtained these from the $g$ tensor's principal values and axes (their Table~3 and Eq.~(8)), together with the canting angle ${\phi^{(i,i')}\approx (-1)^{i}\, 30.67^\circ}$ (right column of Table~1 from their Supplement) upon converting to the global frame $(\hat{a},\hat{b},\hat{c})$ \cite{gtensorfoot}. We note that in Ref.~\onlinecite{woodland23II} the canting angle ${\psi^{(i,i')}\coloneq\sphericalangle(\hat{Z}^{(i,i')},\hat{c})}$ of the $g$ tensor's (third) principal axis was not fitted, but set to $\psi^{(i,i')} \equiv (-1)^i\, 30^{\circ}$, while the three principal values were fitted.

\begin{table}[tb]
\centering
\begin{tabular}{|r|r|r|}
\hline
& Konieczna \textit{et al.} \cite{konieczna25} & Woodland \textit{et al.} \cite{woodland23II} \\\hline
$g_{aa}$ & $+4.000$ & $+4.198$ \\\hline
$g_{bb}$ & $+3.130$ & $+3.322$ \\\hline
$g_{cc}$ & $+5.880$ & $+6.003$ \\\hline
$g_{ca}$ & $-1.720$ & $-1.563$ \\\hline
\end{tabular}
\caption{
Non-vanishing entries $g_{\mu\nu}$ of the $g$ tensor as given in  Eq.~\eqref{eq:g_tensor}. The values have been converted to the global crystallographic reference frame $(\hat{a},\hat{b},\hat{c})$ and rounded to four digits of precision.
}
\label{tab:g_values}
\end{table}

\subsection{Interchain coupling}
\label{sec:model:Interchain_coupling}

We now turn to the interchain couplings which determine the 3D magnetic order. The magnetic chains of \conb{} form an array of acute isosceles triangles in the $ab$ plane, with the bond-length ratio being approximately $7.62\,\text{\r{A}}:5.70\,\text{\r{A}}$ \cite{heid95}, Fig.\ref{fig:crystal_structure_CoNb2O6}(b).

Since the DFT study of Konieczna \textit{et al.} \cite{konieczna25} did not account for interchain couplings, we augment their intrachain Hamiltonian by the simplest plausible interchain piece, namely a nearest-neighbor coupling of Heisenberg form \cite{heid95, weitzel00, kobayashi99, kobayashi00, kobayashi04, sarvezuk11I, sarvezuk11II, thota21}:
\begin{align}
&
\hat{H}_{\mathrm{inter}}^{\text{I}}
=
J_{\mathrm{H},\mathrm{l}}
\sum_{ii',j}
\hat{\vec{S}}_j^{(i,i')} \cdot \hat{\vec{S}}_{j}^{(i,i'+2)}
\nonumber
\\
&
+
\alpha J_{\mathrm{H},\mathrm{l}}
\sum_{ii',j}
\hat{\vec{S}}_j^{(i,i')}
\cdot
\Bigl[
\hat{\vec{S}}_{j}^{(i+1,i'+1)}
+
\hat{\vec{S}}_{j}^{(i-1,i'+1)}
\Bigr]
\label{eq:triangular_interchain_coupling}
\end{align}
where the first and second term couple the chains along the shorter base $\parallel\hat{b}$ and the longer legs $\parallel(\pm\hat{a}\pm\hat{b})/2$ respectively,
and $\alpha$ parameterizes the ratio of the corresponding couplings. While Lee \textit{et al.} \cite{lee10} performed their analysis for a relatively small isosceles distortion, ${\alpha=0.99}$ and ${\alpha=0.90}$, experimental fits suggest smaller values of ${\alpha\approx 0.59}$ \cite{heid95} or ${\alpha\approx 0.63}$ \cite{thota21,weitzel00,kobayashi99, kobayashi00, kobayashi04}. To investigate the influence of the parameter $\alpha$, we will compare the resulting phase diagrams for values ${0.6\leq\alpha\leq 0.99}$ that span a sufficiently large region in parameter space.
Given that neutron diffraction \cite{maartense77, scharf79, heid95} shows that the zero-field low-$T$ ordered state displays antiparallel chain order along the short triangular edge, i.e., an $|{\hdots\uparrow\downarrow\uparrow\downarrow\hdots}\rangle$ pattern along the $b$ axis, we require antiferromagnetic interchain coupling, ${J_{\mathrm{H},\mathrm{l}}>}0$ \cite{heid95, kobayashi99, weitzel00, kobayashi00, kobayashi04, sarvezuk11I, sarvezuk11II, lee10}. The one-dimensional behavior of \conb{} at elevated temperatures implies $J_{\mathrm{H},\mathrm{l}}$ being much smaller than the dominant intrachain couplings. Experimental studies indeed agree that $J_{\mathrm{H},\mathrm{l}}$ is approximately one order of magnitude smaller than the intrachain ones \cite{heid95, kobayashi99, weitzel00, kobayashi00, kobayashi04, sarvezuk11I, sarvezuk11II}, suggesting that ${J_{\mathrm{H},\mathrm{l}} \sim 0.1}$\,meV serves as a rough estimate \cite{J_Hl_note}. Below we will consider values $0.05 \leq J_{\mathrm{H},\mathrm{l}} \leq 0.2$ (all in meV).
The full Hamiltonian modeling \conb{} is obtained by adding Eqs.~\eqref{eq:nearest_neighbor_chain}, \eqref{eq:B_field_and_g_tensor} and \eqref{eq:triangular_interchain_coupling}:
\begin{align}
\hat{H}
=
\hat{H}_{\mathrm{NN}|\mathrm{chains}}
+
\hat{H}_{\mathrm{field}}
+
\hat{H}_{\mathrm{inter}}^{\text{I}}.
\label{eq:model_CoNb2O6__KONIECZNA}
\end{align}

The fitting-based model by Woodland \textit{et al.} \cite{woodland23II} employs a more complicated interchain Hamiltonian:
\begin{align}
&
\hat{H}_{\text{inter}}^{\text{II}}
=
J_1\sum_{ii',j}
\hat{\vec{S}}_j^{(i,i')}\cdot\hat{\vec{S}}_j^{(i,i'+2)}
\notag
\\
&
+
J_1'\sum_{ii',j}
\Bigl[
\hat{\vec{S}}_{2j}^{(i,i')}\cdot\hat{\vec{S}}_{2j+1}^{(i,i'-2)}
+
\hat{\vec{S}}_{2j+1}^{(i,i')}\cdot\hat{\vec{S}}_{2j+2}^{(i,i'+2)}
\Bigr]
\notag
\\
&
+
\sum_{ii',j}
\hat{\vec{S}}_j^{(i,i')\,\mathrm{T}}\cdot\mathbf{W}_j^{(i,i')}\cdot
\Bigl[
\hat{\vec{S}}_j^{(i+1,i'-1)}
+
\hat{\vec{S}}_j^{(i+1,i'+1)}
\Bigr].
\label{eq:H_Interchain_Woodland}
\end{align}
The first two terms correspond to couplings along the shorter base of the isosceles triangles ($b$ direction), with a ferromagnetic coupling $J_1 = -8.2$\,µeV between sites within the same layer (i.e. at the same height $c$) and an antiferromagnetic coupling $J_1' = 40.2$\,µeV connecting nearest-neighbor sites between different layers. The last term represents an antiferromagnetic Ising exchange of $J_2 = 23.1$\,µeV along the longer legs of the interchain triangles in the same layer, which yields the coupling matrix
\begin{align}
\mathbf{W}_{j}^{(i,i')}
=
\begin{pmatrix}
W_{aa} & 0 & -(-1)^{i} W_{ac} \\
0 & 0 & 0 \\
(-1)^{i} W_{ac} & 0 & W_{cc}
\end{pmatrix}
\label{eq:H_Interchain_Woodland_Matrix}
\end{align}
in the global reference frame, with matrix elements
\begin{align}
\begin{cases}
W_{aa} &= -J_2\sin^2\psi = -5.8\,\text{µeV} \\
W_{cc} &= J_2\cos^2\psi = 17.3\,\text{µeV} \\
W_{ac} &= -J_2\sin\psi\cos\psi = -10.0\,\text{µeV}
\end{cases}
\label{eq:H_Interchain_Woodland_Matrix_Elements}
\end{align}
and $\psi=30^\circ$. The model by Woodland \textit{et al.} \cite{woodland23II} is then obtained by adding Eqs.~\eqref{eq:nearest_neighbor_chain}, \eqref{eq:Woodland_NNN_Chains}, \eqref{eq:B_field_and_g_tensor}, and \eqref{eq:H_Interchain_Woodland}:
\begin{align}
\hat{H}
=
\hat{H}_{\mathrm{NN}|\mathrm{chains}}
+
\hat{H}_{\mathrm{NNN}|\mathrm{chains}}
+
\hat{H}_{\mathrm{field}}
+
\hat{H}_{\text{inter}}^{\text{II}}.
\label{eq:model_CoNb2O6__WOODLAND}
\end{align}
We note that the idea of modeling the interchain couplings with Ising rather than Heisenberg interactions also appeared earlier in Refs.~\onlinecite{lee10,cabrera14}, where all chains were coupled using pure Ising interactions.


\subsection{Symmetries}
\label{sec:model:symmetries}

The strong spin-orbit coupling breaks all continuous symmetries, leaving a distinct set of discrete symmetries contained in the space group $Pbcn (60)$ of the columbite structure. The non-trivial elements are
\begin{itemize}
\setlength\itemsep{0.1em}
\item $R_{2b}$ (proper twofold rotation $\parallel \hat b$),
\item $S_{2a}, S_{2c}$ (twofold screw axes $\parallel \hat a,\hat c$),
\item $I$ (inversion at the origin),
\item $G_{ab}$ ($n$ glide $\parallel ab$ plane),
\item $G_{ac}, G_{bc}$ (axial glide planes $\parallel ac, bc$ planes).
\end{itemize}
When constructing the corresponding point group, one obtains the Abelian group $D_{2h}^{14} \equiv \mathbb{Z}_2\times\mathbb{Z}_2\times\mathbb{Z}_2$, with the following non-trivial elements:
\begin{itemize}
\setlength\itemsep{0.1em}
\item $R_{2a},R_{2b},R_{2c}$ (twofold rotations about the crystallographic $a,b,c$ axes),
\item $\sigma_{ab},\sigma_{ac},\sigma_{bc}$ ($ab$, $ac$, $bc$ planes as mirror planes),
\item $I$ (inversion at the origin).
\end{itemize}
All point-group symmetries square to unity and are therefore Ising-like symmetries.


\section{Semiclassical analysis}
\label{sec:swt}

A numerically exact treatment of the frustrated three-dimensional spin models at hand is challenging. As we are primarily interested in low-temperature ordered states and their field evolution, we therefore resort to a semiclassical approach.

\subsection{Formalism}

In the first step, the pseudospins $\vec{S}_j^{(i,i')}$ are treated as classical vectors and parameterized by two angles $(\vartheta_j^{(i,i')},\varphi_j^{(i,i')})$ in spherical coordinates. We use periodic boundary conditions together with a finite magnetic unit cell consisting of $N=N_x\times N_y\times N_z$ copies of the crystallographic unit cell and $p=4 N_x N_y N_z$ spins. The Hamiltonian then becomes a functional of the angles, $\hat{H} \mapsto \mathcal{H}[\vartheta_j^{(i,i')}, \varphi_j^{(i,i')}]$. Minimizing this functional for fixed parameters w.r.t. the $2p$ angles and then choosing the magnetic unit cell which yields the lowest energy, we obtain the classical ground state. In practice, we used all cells from size ${1\times 1\times 1}$ up to size ${4\times 4\times 4}$. From the classical ground state, we obtain the longitudinal and transverse magnetization.

Once the classical ground state is known, we switch to quantum spins of size $S$ and consider fluctuations about the classical state using standard spin-wave theory. To this end, each spin is formally rotated into a local frame,
\begin{align}
\hat{\vec{S}}_{j}^{(i,i'), abc}
=
\mathcal{R}^{j,(i,i')} \hat{\vec{S}}_{j}^{(i,i'), \alpha\beta\gamma}
\end{align}
with the orthogonal matrix
\begin{align}
\mathcal{R}^{j,(i,i')}
&=
\begin{pmatrix}
\cos\vartheta\cos\varphi & -\sin\varphi & \sin\vartheta \cos\varphi
\\
\cos\vartheta\sin\varphi & \cos\varphi & \sin\vartheta\sin\varphi
\\
-\sin\vartheta & 0 & \cos\vartheta
\end{pmatrix}
\end{align}
and the short-hand notation ${(\vartheta,\varphi)\equiv (\vartheta_j^{(i,i')}, \varphi_j^{(i,i')})}$. We then apply a Holstein-Primakoff transformation in the local $(\alpha,\beta,\gamma)$ frame, ${\hat{S}_{j}^{(i,i'),\gamma} = S - \hat{a}_j^{(i,i')\,\dagger}\hat{a}_j^{(i,i')}}$ and ${\hat{S}_j^{(i,i'),+} = \sqrt{2S - \hat{a}_j^{(i,i')\,\dagger}\hat{a}_j^{(i,i')}} \,\, \hat{a}_j^{(i,i')}}$, with bosonic magnon operators $\hat{a}_j^{(i,i')}$. The Hamiltonians \eqref{eq:model_CoNb2O6__KONIECZNA}, \eqref{eq:model_CoNb2O6__WOODLAND} take the form
\begin{align}
\hat{H} \approx \hat{H}_0 + \hat{H}_1 + \hat{H}_2 + \ldots
\end{align}
where $\hat{H}_n$ contains products of $n$ bosonic operators and is of order $\mathcal{O}(S^{2-\frac{n}{2}})$. The term $\hat{H}_0$ is identical to the classical energy functional. By choosing the angles $\{\vartheta,\varphi\}$ such that $\hat{H}_0$ is minimal, it is also ensured that $\hat{H}_1\equiv 0$. Linear spin-wave theory (LSWT) is contained in $\hat{H}_2$, which is bilinear in $\hat{a}_j^{(i,i')\,(\dagger)}$ and describes quantum corrections to leading order.
We switch to Fourier space by
\begin{align}
\hat{a}_{j}^{(i,i')} = \sqrt{\frac{1}{N}}\sum_{\vec{k}\in\mathrm{MBZ}} \mathrm{e}^{\mathrm{i}\vec{k}\cdot\vec{R}_j^{(i,i')}} \hat{a}_{\vec{k}|\lambda}
\end{align}
where $\vec k$ runs over the magnetic Brillouin zone, and the position of each spin must be decomposed into the respective coordinates of the magnetic unit cell $\vec{R}$ and the site location $\vec{r}_{\lambda}$ ($\lambda=1,\ldots,p$) within the magnetic unit cell, $\vec{R}_j^{\,(i,i')} = \vec{R}+\vec{r}_{\lambda}$. The operator $\hat{H}_2$ is then diagonalized by the Bogoliubov transformation
\begin{align}
\hat{a}_{\vec{k}|\lambda}
=
\sum_{\tau=1}^{p} u_{\lambda\tau}(\vec{k})\hat{\gamma}_{\vec{k}|\tau} + v_{\lambda\tau}(-\vec{k})\hat{\gamma}_{-\vec{k}|\tau}^{\dagger}
\end{align}
with coefficients $u_{\lambda\tau}(\vec{k})$ and $v_{\lambda\tau}(-\vec{k})$ such that the new operators $\hat{\gamma}_{\vec{k}|\tau}$, $\hat{\gamma}_{\vec{k}|\tau}^{\dagger}$ are again bosonic. The explicit construction of the bosonic Bogoliubov transformation is described in detail in the literature \cite{mucciolo04,wessel05,delmaestro04}.
The diagonalization finally leads to
\begin{align}
\hat{H}_2
=
\sum_{\vec{k}\in\mathrm{MBZ}}\sum_{\nu=1}^{p} \omega_{\nu}(\vec{k}) \hat{\gamma}_{\vec{k}|\nu}^{\dagger}\hat{\gamma}_{\vec{k}|\nu}
\end{align}
up to a constant, with ${\omega_{\nu}(\vec{k})\geq 0}$ being the magnon frequencies. The Bogoliubov coefficients $u_{\lambda\nu}(\vec{k})$ and $v_{\lambda\nu}(\vec{k})$ are then used to calculate the tensor components $\mathcal{S}^{\mu\nu}(\vec{q},\omega)$ of the dynamical structure factor. At ${T=0}$, the latter are given by
\begin{align}
&
\mathcal{S}^{\mu\nu}(\vec{q},\omega)
=
-\frac{1}{\pi}  \mathfrak{Im} \llangle \hat{S}^{\mu}_{\vec{q}}\hat{S}^{\nu}_{-\vec{q}}\rrangle(\omega)
\nonumber
\\
&=
\sum_{m} \langle 0|\hat{S}^{\mu}_{\vec{q}}|m\rangle\langle m|\hat{S}^{\nu}_{-\vec{q}}|0\rangle
\times
\delta[\omega - (E_m-E_0)]
\label{eq:dynamical_structure_factor}
\end{align}
where $\vec{q}$ is the external momentum, 
and the $\delta$ peaks will be approximated by Gaussians of finite width $\eta$. The single-mode approximation restricts the $|m\rangle$ to be single-magnon states, with energy $E_m$, and $|0\rangle$ is the LSWT ground state, i.e., the vacuum of the $\hat{\gamma}_{\vec{k}|\nu}$ bosons, with energy $E_0$.

\subsection{Field correction}
\label{sec:fieldcorr}

The semiclassical approach does not properly account for ground-state quantum fluctuations. While their effect can be expected to be weak if each chain is deep in its ordered regime, this is not the case for transverse fields near the single-chain critical field. This becomes clear upon considering a single transverse-field Ising chain: The latter can be solved exactly, and the exact critical field differs from that obtained in the semiclassical limit by a factor of two, see Appendix~B.

In order to produce results comparable to experiment, we therefore choose to rescale the magnetic field entering our calculation as follows:
In the Zeeman term \eqref{eq:B_field_and_g_tensor} we rescale the effective field $\mathbf{g}^{(i,i')}_{j} \cdot \vec{B}$ acting on each spin $\hat{\vec{S}}_j^{(i,i')}$ such that the components perpendicular to the easy axis of chain $(i,i')$ are amplified by a factor of two. This ensures the correct transverse critical field in the Ising-chain limit, while leaving the longitudinal field components -- which control interchain order -- untouched.


\section{Ground-state phases}
\label{sec:phases}

As the triangular lattice of chains, i.e., in the $ab$ plane, is frustrated, the interplay between the interchain interactions, spin-orbit coupling, and the external magnetic field drives the system into a variety of interesting phases. To enumerate the phases, we use the convention of Weitzel \textit{et al.} \cite{weitzel00}.

\subsection{Single chain}
\label{sec:phases:single_chain}

To set the stage, we study the physics of an isolated chain. In zero field, each chain $(i,i')$ settles in a state with parallel spins, resulting in ferromagnetic order. The uniform magnetization points along the easy axis $\hat{n}^{(i,i')}$ bisecting the two local Ising axes $\hat{n}_1^{(i,i')},\hat{n}_2^{(i,i')}$. The order breaks a $\mathbb{Z}_2$ symmetry which is \textit{not} a global spin-flip symmetry, as this is broken by the general nearest-neighbor Hamiltonian, but the $c$ glide symmetry \cite{fava20}.

The canting angle $\phi^{(i,i')}$ between the easy axis $\hat{n}^{(i,i')}$ and the $c$ axis requires discussion: The classical computation yields for the DFT-based couplings of Konieczna \textit{et al.} \cite{konieczna25} ${|\phi^{(i,i')}| = 26.46^{\circ}}$, whereas the couplings from
Woodland{}\hfill{}\textit{et}{}\hfill{}\textit{al.}{}\hfill{}\cite{woodland23II}{}\hfill{}result{}\hfill{}in{}\hfill{}${|\phi^{(i,i')}|= 30.00^{\circ}}$,{}\hfill{}to{}\hfill{}be
\begin{figure*}[!t]
\includegraphics[width=0.98\textwidth]{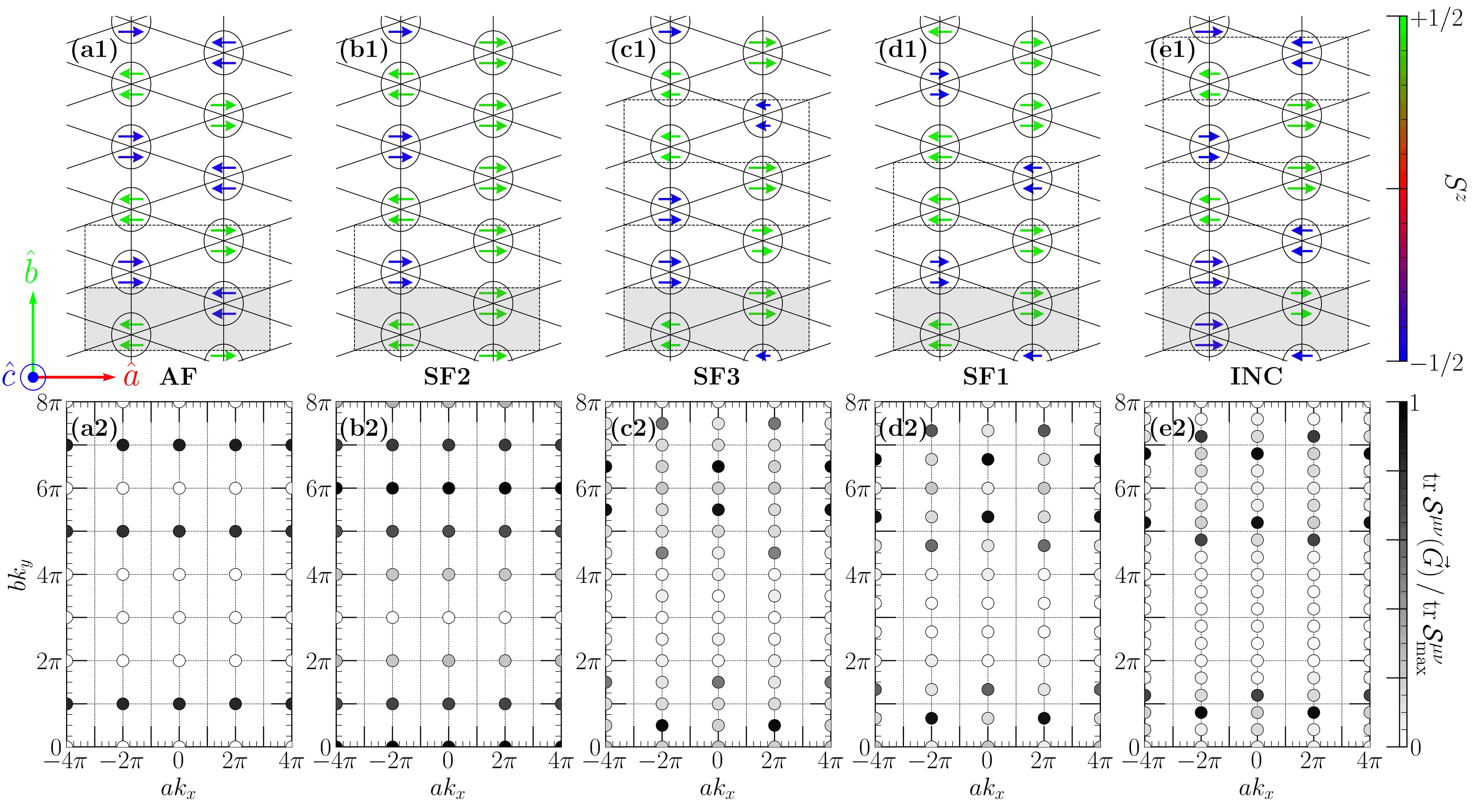}
\caption{
Top: Model ground states for $\vec B$ applied in the $ac$ plane. The crystallographic unit cell (light grey) contains two chains (circles) with two spins (arrows) each. Along each chain, the spins order ferromagnetically. Except for the paramagnetic high-field phase (not shown), all phases require an enlarged magnetic unit cell (dashed). The arrows represent the projection of the spins onto the $ab$ plane; their $c$-axis component is color-coded. Bottom: Corresponding static spin structure factors $\sum_\nu \mathcal{S}^{\nu\nu}(\vec{k})$ evaluated at ${k_z=0}$; the gray scale encodes Bragg-peak intensity. The phase locations are specific to the interchain parameters ${J_{\mathrm{H},\mathrm{l}}=0.2}$\,meV and ${\alpha=0.9}$.
\textbf{(a1-a2):} AF state, with doubled unit cell, realized at zero and small fields. Here ${(B,\Omega_1, \Omega_2)=(0.1\,\mathrm{T},0^{\circ},90^{\circ})}$.
\textbf{(b1-b2):} SF2 state, also with doubled unit cell, reached from AF via a first-order spin-flip transition. Here ${(B,\Omega_1, \Omega_2)=(0.8\,\mathrm{T},0^{\circ},50^{\circ})}$. SF2 and AF can be distinguished by the intensity of the magnetic Bragg peak at ${\vec{k}=\vec{0}}$.
\textbf{(c1-c2):} SF3 state, with quadrupled (${1\times 4\times 1}$) unit cell and additional Bragg peaks, reached via another spin-flip transition. Here ${(B,\Omega_1,\Omega_2)=(0.646\,\mathrm{T}, 0^{\circ},60^{\circ})}$.
\textbf{(d1-d2):} SF1 state, with tripled (${1\times 3\times 1}$) unit cell; this phase is also called FR in other papers \cite{kobayashi99,lee10}. Here ${(B,\Omega_1,\Omega_2)=(0.35\,\mathrm{T}, 0^{\circ},0^{\circ})}$. Six dominant Bragg peaks are at ${(k_x,k_y) = (0,\pm\frac{4\pi}{3b})}$ and ${(k_x,k_y) = (\pm\frac{2\pi}{a},\pm\frac{2\pi}{3b})}$.
\textbf{(e1-e2):} Incommensurate phase INC, reached at elevated fields. The phase displays a dense set of Bragg peaks along the $k_y$ direction \cite{lee10}, most of which have very low intensity such that the pattern looks similar to that of SF1. The figure shows a commensurate approximation with (${1\times 5\times 1}$) unit cell; here ${(B,\Omega_1,\Omega_2)=(0.468\,\mathrm{T},0^{\circ},90^{\circ})}$, and high-intensity peaks are at ${(k_x,k_y) = (0,\pm\frac{6\pi}{5b})}$ and ${(k_x,k_y) = (\pm\frac{2\pi}{a},\pm\frac{4\pi}{5b})}$.
}
\label{fig:ground_states}
\end{figure*}

~\\[-1em]compared with the experimental values listed in Table~\ref{tab:Ising_axis_canting_angle}.

We estimate the quantum corrections to the angle \cite{gallegos24} by performing exact diagonalization for finite $S=1/2$ chains subject to the single-chain Hamiltonian \eqref{eq:nearest_neighbor_chain}, where the canting angle of the $n$-th spin can be obtained by
$|\phi_n| = \arctan(\sqrt{\langle S_n^x\rangle^2 + \langle S_n^y\rangle^2}/ \langle S_n^z\rangle)$.
The expectation value is taken with the ground state of the chain, with a very small field applied to split the Ising degeneracy. Finite-size scaling for chain lengths up to $L=20$ results in ${|\phi_j^{(i,i')}|\approx 27.57^{\circ}}$ for the parameter set of Ref.~\onlinecite{konieczna25}, i.e., quantum corrections to the angle are small.

For a field applied perpendicular to the easy axis, the symmetry-breaking order in the single chain survives up to a critical field $B_{c1}$ where a transition into a field-polarized state occurs; the transition is in the (1+1)-dimensional Ising universality class \cite{fava20}. For the parameters of Refs.~\onlinecite{konieczna25} and \onlinecite{woodland23II} we obtain ${B_{c1}\approx 5.31}$\,T and $4.40$\,T, respectively, with the field correction of Sec.~\ref{sec:fieldcorr} applied. For \conb{} the single-chain critical field has been estimated as $5.25\pm 0.15$\,T \cite{kinross14}.

\subsection{Coupled chains at zero field}

The magnetic interchain couplings induce three-dimensional long-range order. The couplings of Konieczna \textit{et al.} \cite{konieczna25} include antiferromagnetic interchain Heisenberg coupling, $J_{\mathrm{H,\mathrm{l}}} > 0$. Due to the isosceles distortion of the triangular lattice of chains, $\alpha < 1$, the energetic cost for a misaligned spin is higher along the shorter base of the triangle than for the longer legs. The ground state therefore realizes a magnetic unit cell of size ${1\times 2\times 1}$, with antiparallel spins along the $b$ direction, resulting in the antiferromagnetic state AF. The relative

\begin{figure*}[!t]
\centering
\includegraphics[width=\textwidth]{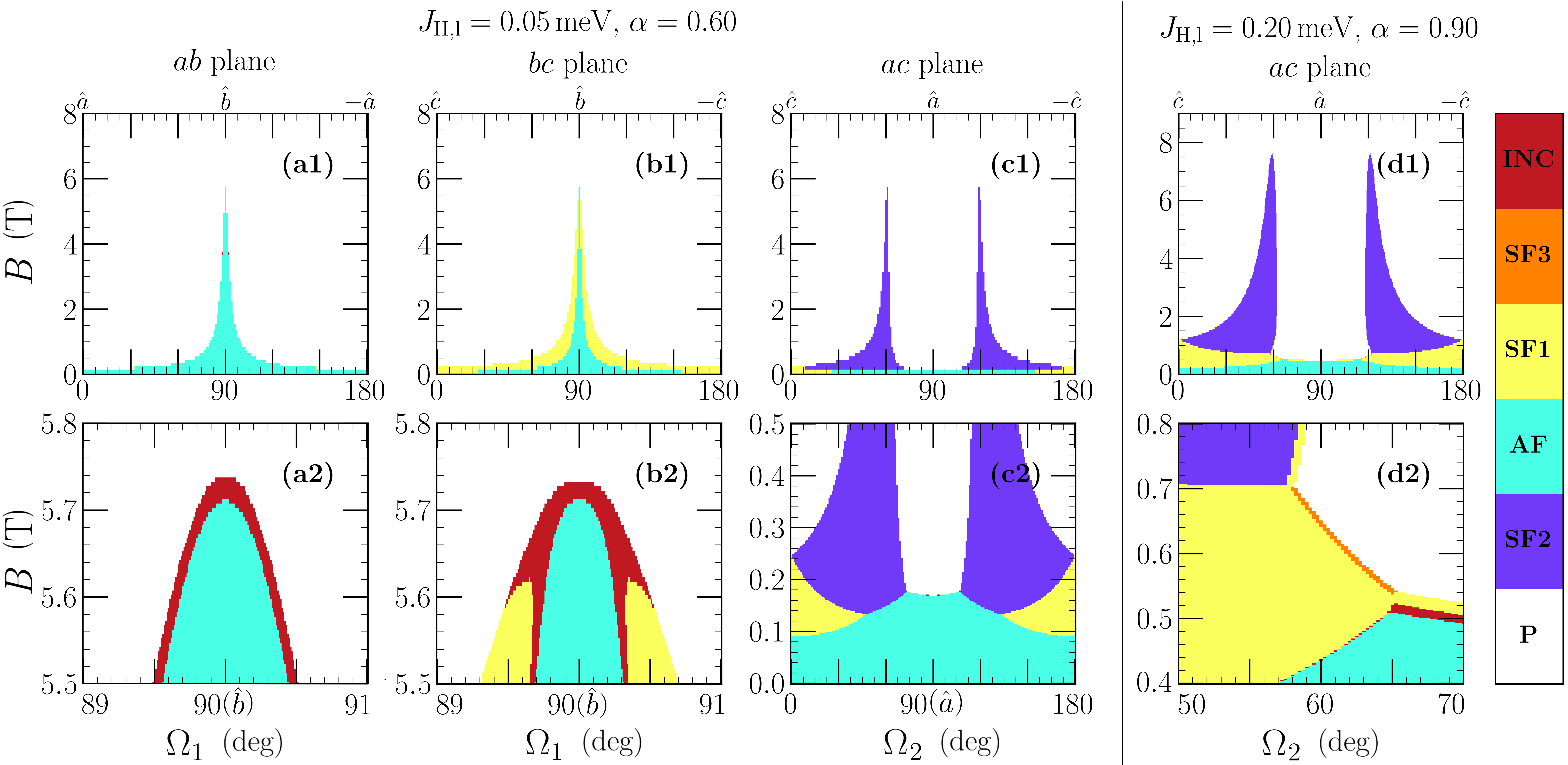}
\caption{
Model phase diagrams as function of field magnitude and direction, with phase labels and colors shown on the right.
Columns \textbf{(a-c)} employ interchain Heisenberg couplings with ${J_{\mathrm{H},\mathrm{l}}=0.05\,\mathrm{meV}}$ and ${\alpha=0.6}$, whereas column \textbf{(d)} uses ${J_{\mathrm{H},\mathrm{l}}=0.2\,\mathrm{meV}}$ and ${\alpha=0.9}$, both with the intrachain couplings by Konieczna \textit{et al.} \cite{konieczna25}.
\textbf{(a):} $\vec{B}$ in the $ab$ plane.
\textbf{(b):} $\vec{B}$ in the $bc$ plane.
\textbf{(c,d):} $\vec{B}$ in the $ac$ plane.
The upper row shows the full angle and field range, while the lower row shows suitable zooms.
The SF2 phase is seen to extend to rather large fields for ${\Omega_2\approx 60^\circ, 120^{\circ}}$; this is because ${\Omega_2=63.54^{\circ},116.46^{\circ}}$ correspond to field directions perpendicular to \textit{one} of the Ising axes, such that in-chain order is particularly stable.
}
\label{fig:pd1}
\end{figure*}

orientation of spins between the two inequivalent chains per crystallographic unit cell is arbitrary, leading to a fourfold degenerate ground state. In Ref.~\onlinecite{lee10}, the state AF is referred to as N\'eel state N2.


As shown in Fig.~\ref{fig:ground_states}(a1) below, all spins are aligned in the $ac$ plane. The component $S^y$ vanishes (as dictated by the local Ising axes) and the other components $S^x,S^z$ alternate such that the ground state has no net magnetization, in agreement with neutron diffraction experiments \cite{morris21,scharf79}. Since the interchain coupling is $\mathrm{SU}(2)$ invariant, the angle $|\phi^{(i,i')}|$ stays the same as for decoupled chains.

For the couplings reported by Woodland \textit{et al.} \cite{woodland23II}, we find that the system also realizes the antiferromagnetic ground state AF.

\subsection{Field-induced phases and phase diagram}

We now turn to an overview of the phases that are realized in an applied field. To parametrize the external field $\vec{B}$, we use spherical coordinates
\begin{align}
\vec{B}/B
&=
 \hat{a}\cos\Omega_1\sin\Omega_2
+\hat{b}\sin\Omega_1
+\hat{c}\cos\Omega_1\cos\Omega_2
\label{eq:Parametrization_of_B}
\end{align}
with angles $\Omega_1\in[0^{\circ},360^{\circ})$ and $\Omega_2\in[0^{\circ},180^{\circ})$; $\Omega_1=90^{\circ}$ then corresponds to a transverse field, $\vec{B}=B\hat{b}$, independent of $\Omega_2$.

In total we find six different phases, each with a magnetic unit cell of size ${1\times N_y \times 1}$ as illustrated in Fig.~\ref{fig:ground_states}:
\begin{itemize}
\item a paramagnetic high-field phase P ($N_y=1$), \\[-6mm]
\item the antiferromagnetic phase AF ($N_y=2$), \\[-6mm]
\item three different spin-flip phases SF1  ($N_y=3$), SF2  ($N_y=2$), SF3  ($N_y=4$), all with ferrimagnetic spin patterns, and \\[-6mm]
\item an incommensurate phase INC (formally $N_y=\infty$), see Sec.~\ref{subsec:INC_Discussion}.
\end{itemize}
These phases have indeed been observed in various experiments \cite{scharf79, cabrera14, heid97, heid95, kobayashi99, kobayashi04, kobayashi00, mitsuda94, hanawa94, morris14, sarkis21, sarvezuk11I, thota21, weitzel00, woodland23I, woodland23II}.

Keeping the intrachain couplings fixed, the extent and location of those phases in the model's $T=0$ phase diagram as function of strength and direction of $\vec B$ depend significantly on the interchain parameters $J_{\mathrm{H},\mathrm{l}}$ and $\alpha$. To illustrate the phase diagram's complexity, we show results for two different sets of $\{J_{\mathrm{H},\mathrm{l}}, \alpha\}$ for the $ab$, $bc$, and $ac$ planes in Fig.~\ref{fig:pd1}. We find narrow strips of INC separating (AF,SF1) and (AF,P) as well as islands of SF3 separating (SF1,P), leading to several triple points. Most transition fields increase with both increasing $J_{\mathrm{H},\mathrm{l}}$ and $\alpha$ which is plausible given that $J_{\mathrm{H},\mathrm{l}}$ and $\alpha J_{\mathrm{H},\mathrm{l}}$ are the interchain couplings stabilizing the ordered states.

A qualitative comparison with \conb{} experiments \cite{heid95,heid97,weitzel00}, performed at temperatures down to 1.5\,K, shows that the sequence of phases agrees for $\vec{B}\parallel\hat{b}$, AF$\to$INC$\to$P, and $\vec{B}\parallel\hat{c}$, AF$\to$INC$\to$SF1$\to$SF2$\to$P. For $\vec{B}\parallel\hat{a}$ we find AF$\to$INC$\to$SF1$\to$P whereas the exper\-iment
\pagebreak
\begin{figure}[!t]
\includegraphics[width=\columnwidth]{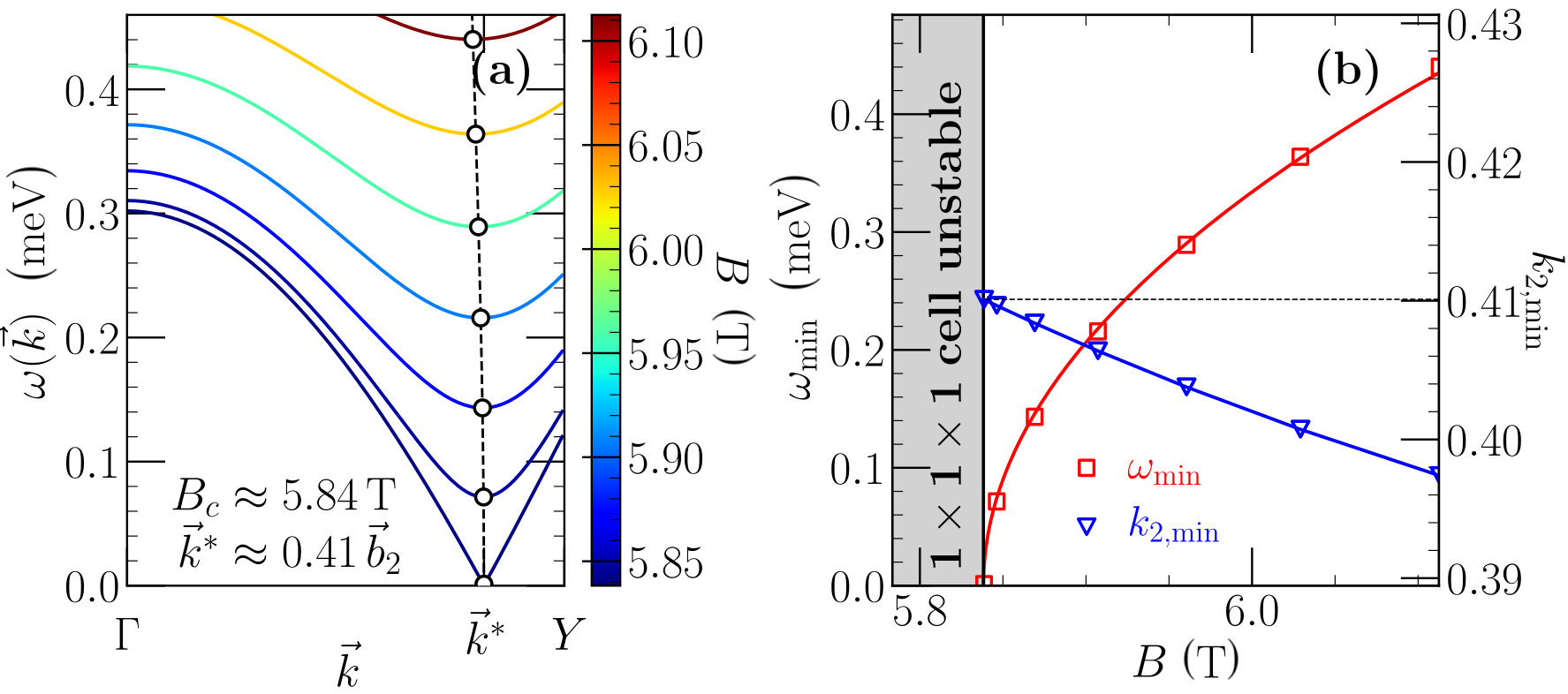}
\caption{
\textbf{(a):} Dispersion of the lowest magnon branch along $\Gamma$-$Y$ in the high-field phase for different ${\vec{B}\parallel\hat{b}}$, computed with the intrachain couplings by Konieczna \textit{et al.} together with ${J_{\mathrm{H},l}=0.05}$\,meV and ${\alpha=0.9}$. The dispersion minimum is at ${\vec{k}=(0,k_{2,\mathrm{min}},0)}$.
\textbf{(b):} Magnon gap $\omega_{\mathrm{min}}$ and corresponding wavevector $k_{2,\mathrm{min}}$ as a function of $B$. The gap closes according to ${\omega_{\mathrm{min}}\propto (B-B_c)^{1/2}}$, with ${B_c\approx 5.84}$\,T. At $B_c$ we have ${k_{2,\mathrm{min}}\approx 0.41}$, signaling the onset of the INC phase.
}
\label{fig:gap_closing}
\end{figure}
\cite{weitzel00} reports AF$\to$INC$\to$SF3$\to$P; our SF3 occurrence is instead restricted to field directions away from the main axes, Fig.~\ref{fig:pd1}(d2).
Parenthetically, we note that the perturbative treatment of interchain couplings in Ref.~\onlinecite{lee10} predicts an intermediate SF1 phase for $\vec{B}\parallel\hat{b}$; this is not observed in our calculation and also appears to be absent from experiments \cite{heid95,heid97,weitzel00,engelhardt}.

From Fig.~\ref{fig:pd1} we see that the ordered phases are particularly stable for transverse field $\vec{B}\parallel\hat{b}$; this is where the external field competes with both intrachain and interchain interactions. This also applies to fields which are perpendicular to \textit{one} of the two easy axes, which explains the existence of SF2 at rather large fields in the $ac$ plane for ${\Omega_2\approx 60^\circ, 120^{\circ}}$, Figs.~\ref{fig:pd1}(c1) and (d1) \cite{weitzel_note}. This even leads to reentrant order in Fig.~\ref{fig:pd1}(d1) where for, e.g., ${\Omega_2=60^\circ}$ the phase sequence is AF$\to$INC$\to$SF1$\to$SF3$\to$P$\to$SF1$\to$SF2$\to$P.
In contrast, for fields $\vec{B}\parallel\hat{a}$ and $\vec{B}\parallel\hat{c}$ (and other ``generic'' directions) the 3D order is destroyed already at much smaller fields; here the external field competes with the interchain coupling only. Given that the in-chain order is ferromagnetic in all ordered phases, most phases and transitions can be understood in terms of a triangular lattice of effective macrospins formed by the individual chains. For transverse field, $\vec{B}\parallel\hat{b}$, these macrospins become soft (i.e. their amplitude vanishes) at the high-field transition, whereas the other transitions can be mainly understood in terms of hard macrospins.

The quantitative comparison of the transition fields shows that detailed agreement between experiment and the present theory cannot be achieved. Considering the transition to the high-field phase for $\vec{B}\parallel\hat{b}$, whose location has been estimated as ${B_c\approx 5.5}$\,T \cite{coldea10} or ${B_c\approx 5.7}$\,T \cite{engelhardt}, suggests that the model with ${J_{\mathrm{H},\mathrm{l}}=0.05}$\,meV and ${\alpha=0.6}$ matches the experiment; this is close to the values suggested earlier in Ref.~\onlinecite{weitzel00}. However, the model calculation yields an incommensurate phase INC whose width comes out far too small. The latter width increases significantly with $J_{\mathrm{H},\mathrm{l}}$ and $\alpha$, however, it is still too small for ${J_{\mathrm{H},\mathrm{l}}=0.2}$, ${\alpha=0.9}$, see also Fig.~\ref{fig:J_alpha_variation} below. This suggests that quantum fluctuations, missing from our semiclassical treatment, stabilize the INC phase against its competitors. For a more detailed comparison and discussion we refer the reader to Sec.~\ref{sec:comp}.

For the parameter set of Woodland \textit{et al.} \cite{woodland23II} we find the same six phases, with quantitative differences concerning the range of phases and location of phase boundaries. To be specific, we employ the intrachain couplings of Konieczna \textit{et al.} \cite{konieczna25}, together with the interchain Heisenberg interaction in Eq.~\eqref{eq:model_CoNb2O6__KONIECZNA}, in the remainder paper, and we relegate the numerical results obtained for the parameters of Ref.~\onlinecite{woodland23II} to Appendix~\ref{app:woodland}.

\subsection{Incommensurate phase}
\label{subsec:INC_Discussion}
At ${T=0}$, the incommensurate phase INC is most
prominent in the close vicinity of the transverse field
direction, ${\vec{B}\parallel\hat{b}}$, where it separates the antiferromagnetic state AF from the paramagnetic high-field phase P, Fig.~\ref{fig:pd1}(a2,b2); it also exists as a narrow strip  for other field directions, Fig.~\ref{fig:pd1}(d2).
In \conb{} experiments, INC also occurs prominently at elevated temperature and zero field \cite{heid95,heid97,weitzel00}, supporting the idea that it is a fluctuation-enhanced phase.

Experimentally, the ordering wavevector of INC is along $\vec{b}_2$ (with $\vec{b}_{1,2,3,}$ the reciprocal-space basis vectors of the orthorhombic structure), ${\Qinc = \qinc \vec{b}_2}$. Its position $\qinc$ varies between 0.37 and 0.48 as function of temperature and magnetic field \cite{heid95,heid97,weitzel00}. In our ${T=0}$ model calculation, we can determine $\qinc$ at the INC$\leftrightarrow$P transition for ${\vec{B}\parallel\hat{b}}$: This transition is continuous, and the excitation spectrum in the high-field P phase gets soft at a field $B_c$ and a wavevector $\vec{k}^{\ast}$ such that ${\Qinc(B_c) = \vec{k}^{\ast}}$. This is illustrated in Fig.~\ref{fig:gap_closing} for a set of model parameters where we find ${\vec{k}^{\ast} = 0.41 \vec{b}_2}$. For explicit computations inside the INC phase we choose the commensurate approximation ${\qinc=2/5}$ which allows us to work with a magnetic unit cell of size ${1\times 5\times 1}$ with ${p=20}$ spins. According to Refs.~\onlinecite{coldea10,lee10}, the magnitude $\qinc$ of the incommensurate ordering wavevector slightly increases in a continuous fashion upon decreasing $B$ which will not be captured by this approximation. We note that the spin configuration of INC displays a non-trivial staggering of both the $a$ and $c$ components of the spins, Fig.~\ref{fig:ground_states}(e1).

It is worth mentioning that the occurrence of incommensurate order in \conb{} was discussed early on \cite{heid95,heid97,weitzel00} in terms of soliton condensation within the axial next-nearest-neighbor Ising (ANNNI) model \cite{bak80}, i.e., an Ising model with competing frustrated interactions. However, such a model is insufficient to capture the full complexity of the phase diagram of \conb{}.

\subsection{Variation of interchain couplings}
\label{subsec:J_alpha_Variation}
\begin{figure*}[!t]
\centering
\includegraphics[width=0.85\textwidth]{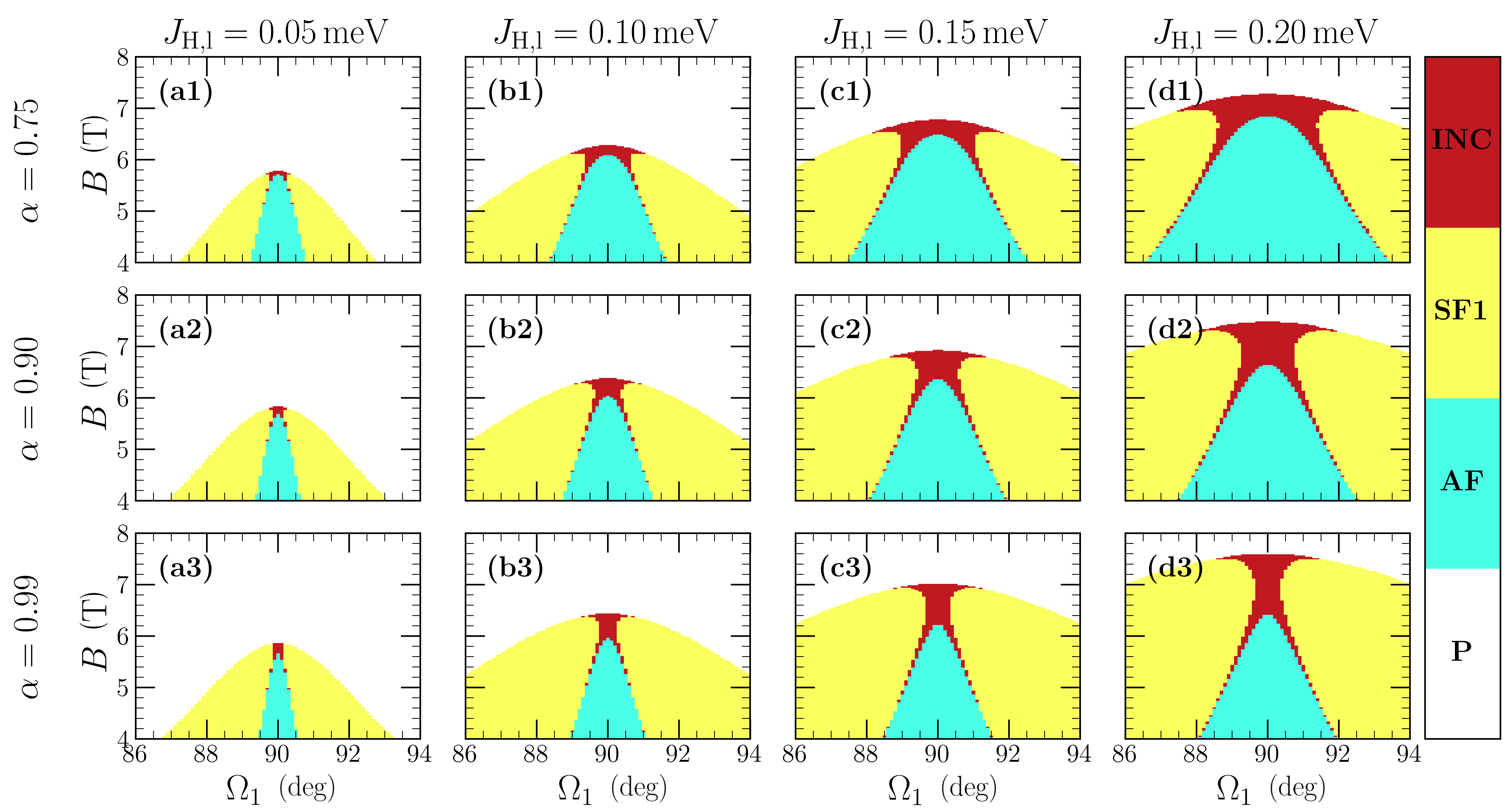}
\caption{
Model phase diagrams as function of strength and direction of the applied field, focusing on nearly transverse fields in the $bc$ plane near the transition into the high-field phase; $\Omega_1=90^{\circ}$ corresponds to $\vec{B}\parallel\hat{b}$, i.e., transverse field. Results are shown for various choices of the interchain couplings $\alpha$ and $J_{\mathrm{H},\mathrm{l}}$; the intrachain couplings are those of Ref.~\onlinecite{konieczna25}.
}
\label{fig:J_alpha_variation}
\end{figure*}

As noted earlier, the parameters of interchain couplings in {\conb} are not known to good accuracy, but strongly influence the extend and shape of the low-$T$ ordered phases. To illustrate this, we now showcase several properties as function of the interchain Heisenberg coupling $J_{\mathrm{H},\mathrm{l}}$ and the isosceles distortion $\alpha$, keeping the intrachain couplings fixed to the values of Ref.~\onlinecite{konieczna25}.

Fig.~\ref{fig:J_alpha_variation} shows a piece of the phase diagram near the transverse field direction, i.e., for fields in the $bc$ plane with a tilting away from $\hat b$ of up to $4^{\circ}$. Three trends are obvious: (i) The phase diagram depends very sensitively on the field direction, with $2^\circ$ tilt shifting phase boundaries by several Tesla and even changes the sequence of phases. (ii) Increasing values of $J_{\mathrm{H},\mathrm{l}}$ increase the field range of the symmetry-broken phases, as all those phases are stabilized by the interchain couplings. (iii) Smaller values of $\alpha$ enlarge the INC phases along the $\Omega_1$ axis. The choice of $\alpha$ also influences the overall shape of the INC region, as is clear from the right-most column of Fig.~\ref{fig:J_alpha_variation}. This can be rationalized by noting that smaller $\alpha$ implies less interchain frustration, which stabilizes AF against more complicated ordering patterns.

\begin{figure}[!t]
\centering
\includegraphics[width=0.91\columnwidth]{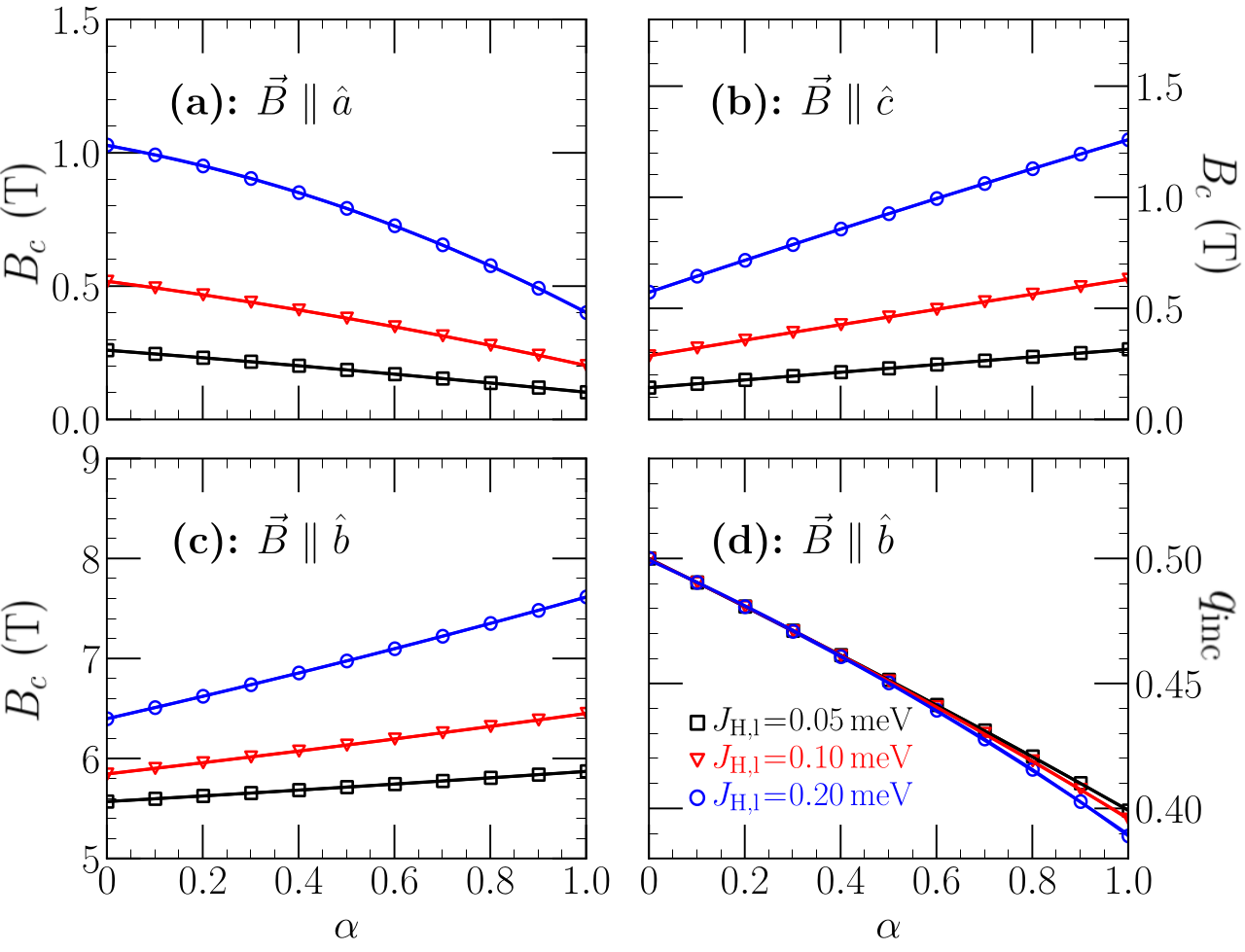}
\caption{
\textbf{(a,b,c):} Transition fields into the high-field phase for different interchain coupling parameters $J_{\mathrm{H},\mathrm{l}}$ and $\alpha$; the intrachain couplings are those of Ref.~\onlinecite{konieczna25}.
\textbf{(a)} SF1$\leftrightarrow$P for $\vec B\parallel \hat{a}$,
\textbf{(b)} SF2$\leftrightarrow$P for $\vec B\parallel \hat{c}$,
\textbf{(c)} INC$\leftrightarrow$P for $\vec B\parallel \hat{b}$.
Panel \textbf{(d)} shows the value of $\qinc$ at the INC$\leftrightarrow$P transition.
}
\label{fig:BcJH}
\end{figure}

Figs.~\ref{fig:BcJH}(a,b,c) shows the variation of the transition fields into the high-field phase for the crystallographic high-symmetry directions. As above, larger $J_{\mathrm{H},\mathrm{l}}$ tend to stabilize the ordered phases. The trend with $\alpha$ is more involved: Larger $\alpha$ implies both more frustration and larger total interchain coupling, the latter in general stabilizing order. Notably, the smaller interchain couplings
$\alpha J_{\mathrm{H},\mathrm{l}}$ are partially frustrated in all phases, but are most frustrated in the canted phases for $\vec B\parallel \hat{a}$, explaining the trend observed in Fig.~\ref{fig:BcJH}(a).

Finally, Fig.~\ref{fig:BcJH}(d) shows the variation of the INC ordering wavevector at $B_c$ for ${\vec{B}\parallel\hat{b}}$. $\qinc$ approaches $1/2$ for $\alpha\to0$, i.e. INC disappears for ${\alpha=0}$: This corresponds to the unfrustrated limit without soliton condensation. Conversely, increasing frustration $\alpha$ increases the soliton density and hence decreases $\qinc$, with a weak dependence on $J_{\mathrm{H},\mathrm{l}}$. We recall that our computation does not capture the variation of $\qinc$ inside the INC phase.

\begin{figure*}[!t]
\centering
\includegraphics[width=0.9\textwidth]{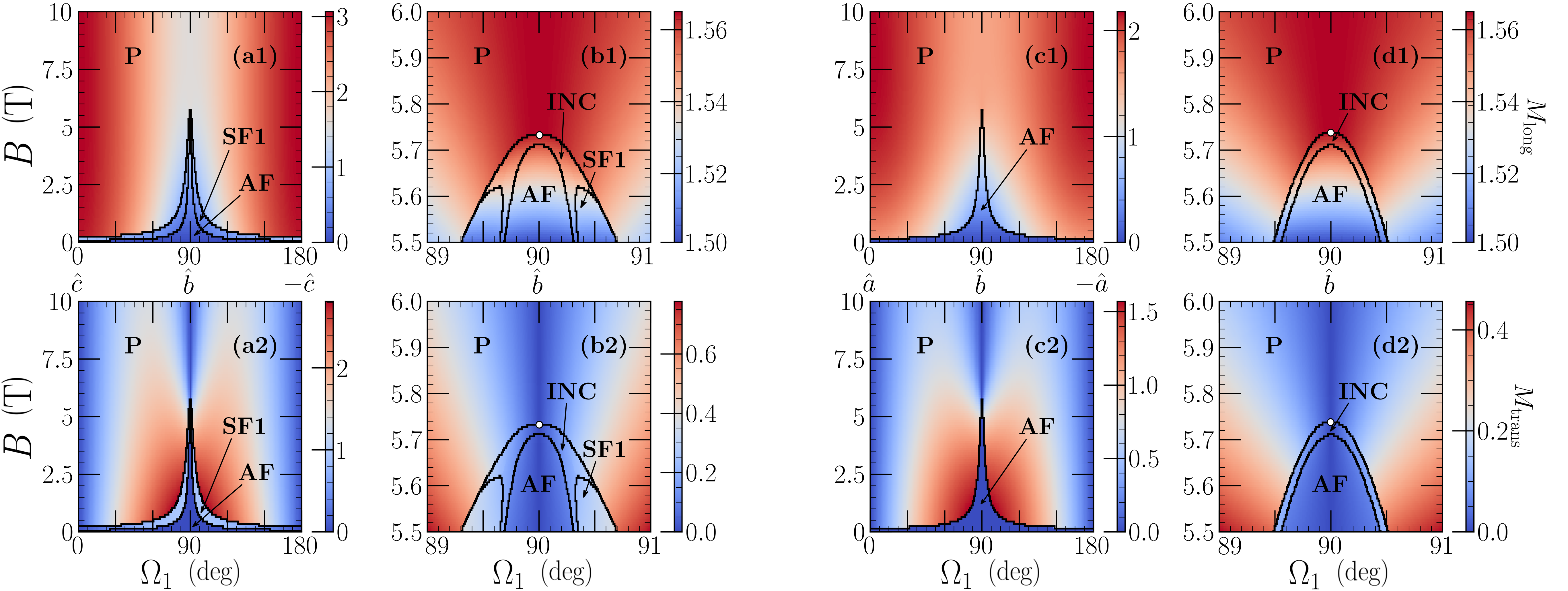}
\caption{
Longitudinal (top) and transverse (bottom) components of the magnetization (color-coded) as a function of field magnitude and angle along two selected cuts, for model parameters ${J_{\mathrm{H},\mathrm{l}}=0.05}$\,meV and ${\alpha=0.6}$.
\textbf{(a,b):} ${\Omega_2=0^{\circ}}$ ($bc$ plane).
\textbf{(c,d):} ${\Omega_2=90^{\circ}}$ ($ab$ plane).
Panels \textbf{(a,c)} show the full angle range, while panels \textbf{(b,d)} contain a zoom close to the transverse direction, ${\Omega_1=90^\circ}$.
All phase transitions are of first order except for ${\mathrm{INC}\leftrightarrow\mathrm{P}}$ which is continuous for transverse field ${\vec{B}\parallel\hat{b}}$ (white circle) but becomes weakly first order if $\vec{B}$ is tilted away from the $b$ axis.
}
\label{fig:Magnetization_Plot}
\end{figure*}

%
%
%
\section{Magnetization}
\label{sec:mag}

Using the classical ground state we can compute the field evolution of the total magnetization (per spin),
\begin{align}
\vec{M} = \frac{1}{N_s}\sum_{ii',j} \langle\mathbf{g}_{j}^{(i,i')}\vec{S}_{j}^{(i,i')}\rangle
=:
\langle \mathbf{g}\vec{S}\rangle
\label{eq:magnetization}
\end{align}
where $N_s$ is the total number of spins and $\langle \mathbf{g}\vec{S}\rangle$ is a short-hand to recall that the quantity $\vec{M}$ incorporates the anisotropic $g$ tensor. $\vec{M}$ can be decomposed into its components parallel and perpendicular to the applied field $\vec B$, $\vec{M}_{\mathrm{long}}$ and $\vec{M}_{\mathrm{trans}}$, respectively. Due to strong spin-orbit coupling and low symmetry, $\vec{M}_{\mathrm{trans}}$ is non-zero (and often sizable) unless $\vec B$ is parallel to one of the main axes $\hat a, \hat b, \hat c$.

In Fig.~\ref{fig:Magnetization_Plot}, we show ${T=0}$ results for both the longitudinal and the transverse magnetization as function of field magnitude and angles in the $bc$ and $ab$ planes, for interchain couplings with ${J_{\mathrm{H},\mathrm{l}}=0.05}$\,meV and ${\alpha=0.6}$. Not surprisingly, the longitudinal magnetization is smallest for transverse fields, ${\vec B\parallel \hat b}$, and is smaller in the symmetry-broken phases as compared to the high-field phase P. The transverse magnetization is sizable for intermediate field directions, i.e., away from the main axis, in particular in the high-field phase P, and only decays to zero as $B\to\infty$. In fact, the transverse magnetization turns out to be a very sensitive probe of the phase transitions, as its value varies strongly between the different phases as function of field angle (at fixed field magnitude), in contrast to the longitudinal magnetization.

\begin{figure}[!b]
\centering
\includegraphics[width=0.8\columnwidth]{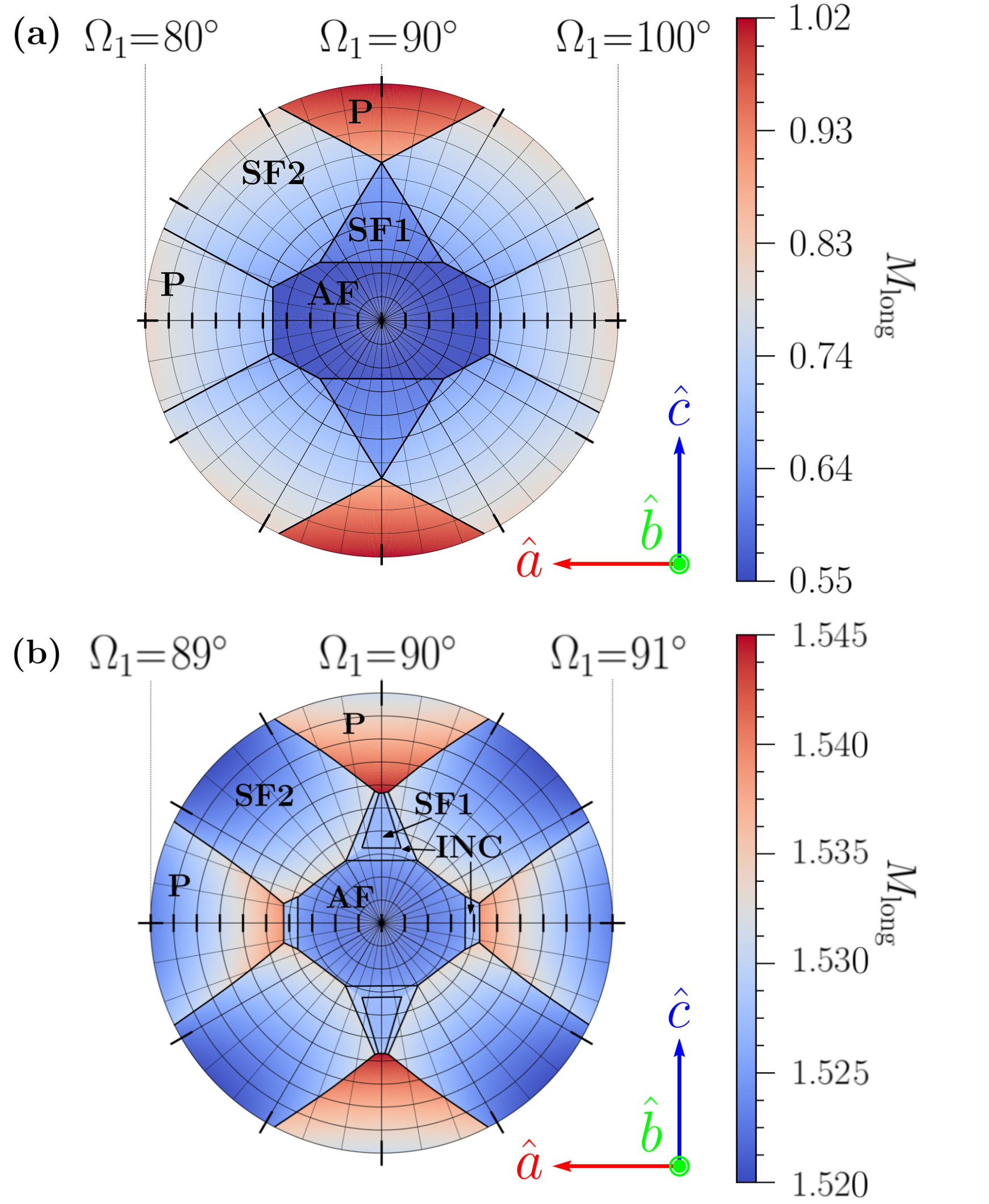}
\caption{
Longitudinal magnetization (color-coded) as function of field direction as in Fig.~\ref{fig:Magnetization_Plot}, now covering a spherical segment near ${\vec{B}\parallel\hat{b}}$.
\textbf{(a):} ${B=2}$\,T, for a range of $10^\circ$ tilt angle.
\textbf{(b):} ${B=5.57}$\,T, for a range of $1^\circ$ tilt angle.
Note that these images obey the point-group symmetries.
}
\label{fig:sphere1}
\end{figure}

\begin{figure*}[!t]
\centering
\includegraphics[width=0.98\textwidth]{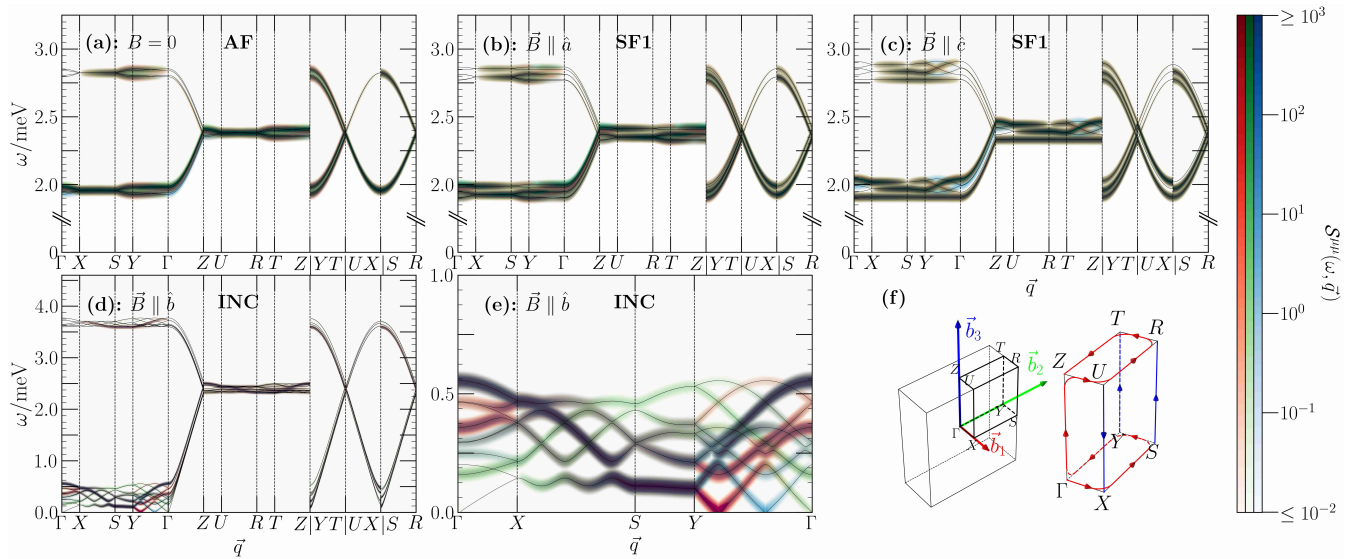}
\caption{
Dynamical structure factor $\mathcal{S}^{\mu\nu}(\vec{q},\omega)$ for selected phases and fields, computed for the intrachain couplings of \cite{konieczna25} and ${J_{\mathrm{H},\mathrm{l}}=0.05}$\,meV and ${\alpha=0.6}$. The $\delta$ peaks were broadened into Gaussians with a width of ${\eta = 0.01}$\,meV; the intensity of the diagonal components $\mathcal{S}^{\mu\mu}$ is shown in a logarithmic color scale.
\textbf{(a):} AF for ${B=0}$.
\textbf{(b):} SF1 for ${\vec{B}\parallel\hat{a}}$ and ${B=0.170}$\,T.
\textbf{(c):} SF1 for ${\vec{B}\parallel\hat{c}}$ and ${B=0.246}$\,T.
\textbf{(d,e):} INC for ${\vec{B}\parallel\hat{b}}$ and ${B=5.72}$\,T, with (e) showing at low-energy zoom.
\textbf{(f):} First Brillouin zone for the orthorhombic crystal system, together the path used to plot the excitation spectra \cite{setyawan10}; note that this path consists of a connected piece (red) and disconnected ones (blue).
}
\label{fig:dynsf1}
\end{figure*}

Most phase transitions can be identified as being first order, with jumps in the both the longitudinal and transverse magnetization, with the exception of INC$\leftrightarrow$P for ${\vec B\parallel \hat b}$.
This implies that field-scan experiments, i.e., the variation of either the magnitude or the direction of $\vec{B}$ at fixed low $T$, will encounter hysteresis effects. Also, transitions involving in/out of the INC phase can be expected to be accompanied by significant dissipation as well as freezing effects, due to the pinning of Ising domain walls.

The sensitivity to field rotation is further illustrated in Fig.~\ref{fig:sphere1} which shows $\vec{M}_{\mathrm{long}}$ for fixed $|\vec{B}|$ as function of field direction. For fields slightly below ${B_c\approx 5.74}$\,T, tilting the field by $1^\circ$ or less from the transverse direction can access five different phases, Fig.~\ref{fig:sphere1}(b).


\section{Excitation spectra}
\label{sec:exc}

We calculate the dynamical structure factor $\mathcal{S}^{\mu\nu}(\vec{q},\omega)$ using LSWT in single-mode approximation, Eq.~\eqref{eq:dynamical_structure_factor}. Sample results are shown in Fig.~\ref{fig:dynsf1} for different magnetic fields and the model parameters of Ref.~\onlinecite{konieczna25}, with interchain couplings ${J_{\mathrm{H},\mathrm{l}}=0.05\,\mathrm{meV}0}$ and ${\alpha=0.6}$. Specifically, we show the three diagonal components of the tensor, $\mathcal{S}^{xx}$, $\mathcal{S}^{yy}$, $\mathcal{S}^{zz}$, simultaneously color-coded in red, green and blue with subtractive color mixing.
When interpreting the spin-wave results, we have to keep in mind that LSWT is not able to reproduce spectral features typical for one space dimension, such as spinons and their longer-range bound states. Such features are known to be relevant for {\conb} at elevated energies \cite{coldea10,morris14,morris21,amelin20I,amelin20II,woodland23I,woodland23II,kjall11}. In contrast, spin-wave theory correctly captures low-energy excitations of 3D ordered states as well as excitations of the high-field state where fractionalization plays little role.

Fig.~\ref{fig:dynsf1}(a) shows the zero-field result for the AF phase while Figs.~\ref{fig:dynsf1}(b,c) display the spectrum in SF1 for small $\vec B$ along $\hat a$ and $\hat c$, respectively. In these cases the spectrum is gapped, with a small bandwidth; the strongest dispersion is for momentum variations parallel along $\vec b_3$ (e.g. $\Gamma$--$Z$ and $Y$--$T$), i.e., along the chain direction. The two main sets of bands observed at ${\vec q=\Gamma}$ near 1.9 and 2.8\,meV arise from the two chain sites per crystallographic cell, with ${\sim 1}$\,meV being the characteristic scale of intrachain couplings; further splitting comes from interchain interactions. The unit-cell tripling of SF1 induces additional spin-wave branches compared to AF.

The situation is different for elevated fields $\vec B\parallel\hat b$ where the excitation gap closes at the continuous P$\leftrightarrow$INC transition. Figs.~\ref{fig:dynsf1}(d,e) show the spectrum in the INC phase near this transition. This spectrum displays sizable dispersion also perpendicular to the chain direction and features a gapless phason mode, with large intensity located at the ordering wavevector $\Qinc$ of INC; this mode corresponds to a sliding mode of the incommensurate structure \cite{lee10,chaikin_lubensky}. We note that our computation is performed in a period-5 approximation to INC, such that the phason mode has a small gap.

\begin{figure*}[!t]
\centering
\includegraphics[width=\textwidth]{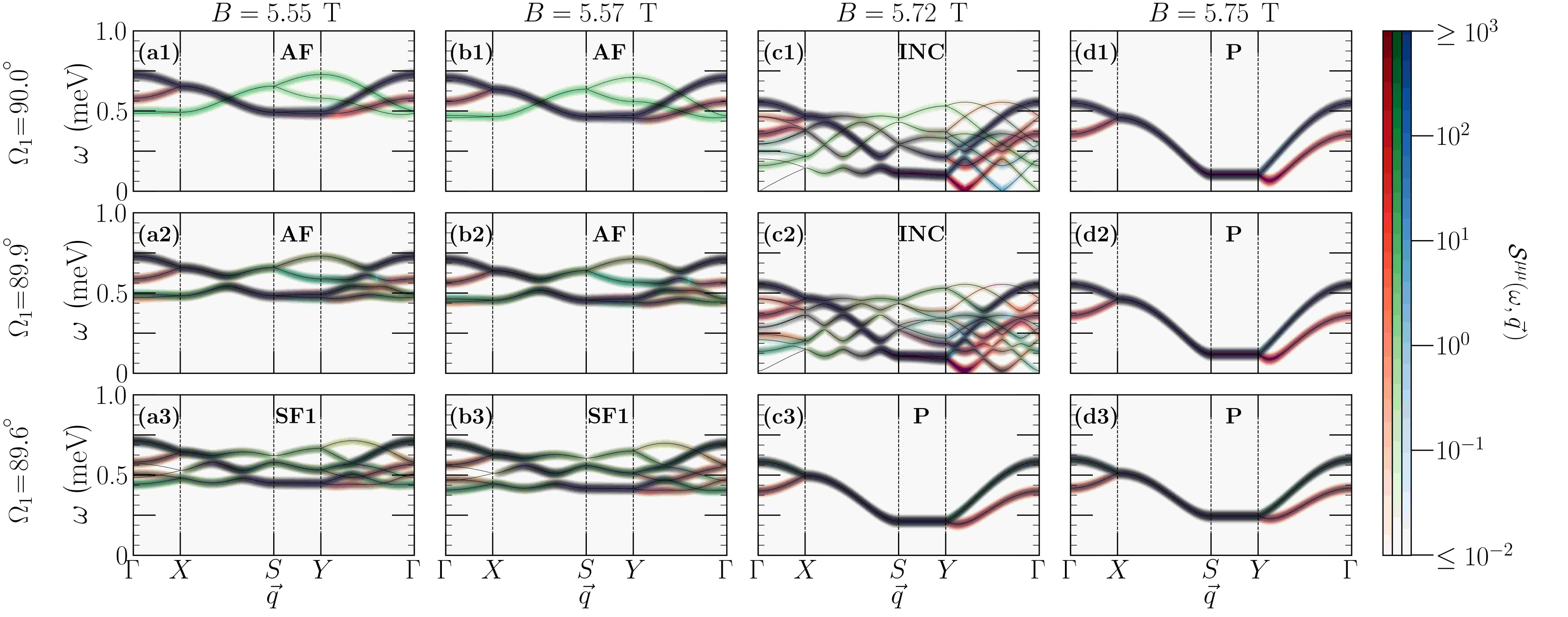}
\caption{
Low-energy part of the dynamical structure factor as in Fig.~\ref{fig:dynsf1}, now focusing on transverse fields ${\vec{B}\parallel\hat{b}}$ (top row) and small tilt angles in the $bc$ plane away from it (lower two rows, ${\Omega_1=89.9^\circ, 89.6^\circ}$), covering a field range ${5.55\leq B \leq 5.75}$ (in T) near the INC$\leftrightarrow$P transition. Phases are marked in the individual panels.
}
\label{fig:dynsf2}
\end{figure*}

A closer inspection of the spectrum near the INC$\leftrightarrow$P transition is in Fig.~\ref{fig:dynsf2}. It shows the dynamical structure factor for different $B$ both above and below the critical field strength $B_c$, for ${\vec B\parallel\hat b}$ and with small tilting into the $bc$ plane. In both AF and SF1 phases, increasing $|\vec{B}|$ decreases the gap and increases the dispersion bandwidth. The gapless phason mode of INC, panels (c1) and (c2), and its precursor in P, panels (d1) and (d2), are nicely seen.
Most significantly, the low-energy part of the spectrum changes drastically upon small tilting away from the $b$ axis: Already for a tilting of $0.4^\circ$ the mode softening of the INC$\leftrightarrow$P transition is no longer visible.


\section{Comparison to experiment}
\label{sec:comp}

\subsection{Phase diagram and transition fields}

The zero-temperature phase diagram obtained from our semiclassical study is largely in qualitative agreement with experimental data, the latter taken at temperatures down to 1.5\,K \cite{heid95,heid97,weitzel00} except for the transverse direction where data down to 0.1\,K \cite{coldea10,engelhardt} are available.
The agreement is semi-quantitative for many features (with the field correction of Sec.~\ref{sec:fieldcorr} taken into account), however, the regions of INC and SF3 phases are too small (in both field magnitude and field angle) in the semiclassical theory compared to experiment. Notably, the experimental data show that the INC phase becomes narrower with decreasing temperature, suggesting that INC gets stabilized by both thermal and quantum fluctuations w.r.t. the neighboring smaller-unit-cell phases.

For a concrete comparison we consider the transition fields to the high-field phase in the (extrapolated) low-$T$ limit; those are for \conb{} for the three crystallographic directions ${B_c^a \approx 0.15}$\,T \cite{heid97,weitzel00} , ${B_c^b \approx 5.7}$\,T \cite{engelhardt}, ${B_c^c \approx 0.4}$\,T \cite{heid95,weitzel00}.
Our theoretical results for the intrachain parameter set of Ref.~\onlinecite{konieczna25} is in Fig.~\ref{fig:BcJH}. This suggests a Heisenberg interchain coupling with ${J_{\mathrm{H},\mathrm{l}}=0.05 \ldots 0.1}$\,meV and ${\alpha=0.6\ldots 0.75}$. As noted above, for those parameters the INC phase is much narrower than in experiment. This could be counter-acted by choosing larger $J_{\mathrm{H},\mathrm{l}}$ and $\alpha$ closer to unity, but such a choice would result in $B_c$ being too large. We conclude that the present analysis is insufficient to fully nail down a parameter set for \conb.
The quantitative disagreement may have multiple sources: (i) Limitations of the effective spin-$1/2$ model: the phase diagram might be influenced by the mixing of excited CEF levels. (ii) Quantum fluctuations: the stability regions of some of the phases, INC and SF3 in particular, might be significantly larger with fluctuations included. Both issues constitute interesting subjects of further studies, but are beyond the scope of this paper.

As noted earlier, our computations predict the spin-flip phase SF2 to extend to rather high fields for intermediate field directions in the $ac$ plane, Figs.~\ref{fig:pd1}(c1,d1). The corresponding trend is visible in Fig.~1 of Ref.~\onlinecite{weitzel00}, but since the feature is narrow in field angle, it has not been properly probed. We therefore suggest further experiments with rotating field to map out this feature.

\subsection{Glassy behavior and freezing}

Measurements of the heat capacity \cite{hanawa94, liang15} and the AC susceptibility \cite{sarkis21} observed freezing effects for transverse fields and at low temperature in and near the INC phase. In the literature, these freezing effects are often referred to as ``glassy'' behavior.

In the INC phase below $B_c$, i.e., in the field range of ${4\ldots 5}\,$T, freezing can be plausibly associated with Ising domain walls. There, the incommensurate order is formed by hard macrospins of homogeneously ordered chains, such that changing the orientation of one of these macrospins involves flipping an entire chain and is thus associated with a large energy barrier. We recall that the ordering wavevector $\Qinc$ depends on temperature, such that cooling is accompanied by a re-arrangement of the Ising pattern. Consistent with the above argument, freezing is weak or absent near $B_c$ because the macrospins get soft and the barriers diminish. It is, however, not obvious why freezing is also seen significantly above $B_c$, i.e. around 7\,T \cite{liang15}. It is open whether this is a precursor effect of INC or, alternatively, whether nuclear spins or quenched disorder, such as defect pinning or a spatially fluctuating $g$ tensor, play a role. Measurements under slightly tilted fields and in differently prepared samples might help to investigate this further.


\section{Summary}
\label{sec:summ}

Using a three-dimensional spin model for \conb, we have mapped out low-temperature magnetic phases and transitions for arbitrary direction of applied magnetic field. The interplay of strong spin-orbit coupling and interchain frustration produces a rich phase diagram, with extreme sensitivity to the direction of the applied field. In particular, the phase transition into the high-field phase, which is continuous for transverse fields, is rendered first order in case of field misalignment; for fields tilted by $0.5^\circ$ or more, even the phase sequence changes.

Our comparison of different parameter sets enabled us to narrow down the range of interchain coupling parameters. We suggest to employ uniaxial pressure along $\hat{a}$ or $\hat{b}$ to tune the interchain coupling and its degree of frustration; we predict that the critical fields as well as the angle range of phases near the critical fields will significantly change under uniaxial pressure, Fig.~\ref{fig:J_alpha_variation}.

The comparison with experiment suggests that the incommensurate phase INC is stabilized by fluctuations, both thermal and quantum, w.r.t. their neighboring phases (the same applies to the phase SF3); this fluctuation enhancement is an interesting subject of further theory work. On the experimental front, more detailed low-temperature measurements near the transverse-field INC--P transition would be helpful to expose its quantum critical behavior \cite{book_sachdev,rop18} and the possible role of quenched disorder. Clearly, crystals with small mosaicity are needed, given the very delicate angle dependence of the field-induced phases.


\acknowledgments

We thank P. M. C\^onsoli, A. Engelhardt, R. K. Kaul, and C. Pfleiderer for illuminating discussions and collaborations on related work, as well as R. Coldea and L. Woodland for clarifications concerning Ref.~\onlinecite{woodland23II}.
This work was funded by the Deutsche Forschungsgemeinschaft (DFG) through SFB 1143 (project id 247310070) and the W\"urzburg-Dresden Cluster of Excellence on Complexity, Topology and Dynamics in Quantum Matter -- \textit{ctd.qmat} (EXC 2147, project id 390858490).

\section*{Data Availability}
The data that support the findings of this article are openly available \cite{data_archive}.

\appendix

\section{Local frames and coupling matrices}
\label{app:coupling_matrices}

The DFT calculation of Ref.~\onlinecite{konieczna25} employs a local reference frame $(\hat{x}^{(i,i')},\hat{y}^{(i,i')},\hat{z}^{(i,i')})$ for each chain $(i,i')$, obtained by taking the Co--O bonds of the octahedra and applying symmetric L\"owdin orthogonalization \cite{lowdin50,lowdin70}. This local coordinate system is given by
\begin{align}
\begin{cases}
\hat{x}^{(i,i')} &= \frac{1}{\sqrt{3}}\hat{a} + \frac{1}{\sqrt{2}}\hat{b} - (-1)^i\frac{1}{\sqrt{6}}\hat{c} \\
\hat{y}^{(i,i')} &= \frac{1}{\sqrt{3}}\hat{a}  + (-1)^i\frac{2}{\sqrt{6}}\hat{c} \\
\hat{z}^{(i,i')} &= \frac{1}{\sqrt{3}}\hat{a} - \frac{1}{\sqrt{2}}\hat{b} - (-1)^i\frac{1}{\sqrt{6}}\hat{c}
\end{cases}
\label{eq:local_frame}
\end{align}
in terms of the global crystallographic axes. In this local frame, the authors introduce and compute nearest-neighbor diagonal couplings of Heisenberg ($J_{\mathrm{H}}$) and Kitaev ($K_y$, $K_z$) type as well as symmetric off-diagonal couplings $\Gamma_{xy}$, $\Gamma_{xz}$, $\Gamma_{yz}$ \cite{koni_note}. Those couplings can be converted into the global frame used in the present paper by applying the orthogonal transformation
\begin{align}
\mathcal{B}_{j}^{(i,i')}
\coloneq
\mathcal{R}
\sigma_{j}^{(i,i')}
=
\begin{pmatrix}
\frac{1}{\sqrt{3}} & \frac{(-1)^j}{\sqrt{2}} & -\frac{(-1)^i}{\sqrt{6}} \\
\frac{(-1)^j}{\sqrt{3}} & 0 & \frac{2(-1)^{i+j}}{\sqrt{6}} \\
\frac{(-1)^i}{\sqrt{3}} & -\frac{(-1)^{i+j}}{\sqrt{2}} & -\frac{1}{\sqrt{6}}
\end{pmatrix}
\end{align}
consisting of a rotation, $\mathcal{R}$, and $\sigma_{j}^{(i,i')}$ which accounts for sign flips according to the glide symmetries. The couplings then transform according to
\begin{align}
&
\mathcal{B}_{j}^{(i,i')\,\mathrm{T}}
\begin{pmatrix}
J_{\mathrm{H}} & \Gamma_{xy} & \Gamma_{xz} \\
\Gamma_{xy} & J_{\mathrm{H}} + K_y & \Gamma_{yz} \\
\Gamma_{xz} & \Gamma_{yz} & J_{\mathrm{H}} + K_z
\end{pmatrix}
\mathcal{B}_{j}^{(i,i')}
\nonumber
\\
&=
\begin{pmatrix}
M_{aa} & (-1)^j M_{ab} & (-1)^i M_{ac} \\
(-1)^j M_{ab} & M_{bb} & (-1)^{i+j} M_{bc} \\
(-1)^i M_{ac} & (-1)^{i+j} M_{bc} & M_{cc}
\end{pmatrix}
\end{align}
where the last form is the one in Eq.~\eqref{eq:Nearest_Neighbor_Matrix} of the main text.
Explicitly, we obtain
\begin{align}
M_{aa} &= (3J_{\mathrm{H}} + K_y + K_z + 2\Gamma_{xy} + 2\Gamma_{xz} + 2\Gamma_{yz}) / 3, \notag\\
M_{bb} &= (2J_{\mathrm{H}} + K_z - 2\Gamma_{xz}) / 2, \notag\\
M_{cc} &= (6J_{\mathrm{H}} + 4K_y + K_z - 4\Gamma_{xy} + 2\Gamma_{xz} - 4\Gamma_{yz}) / 6, \notag\\
M_{ab} &= (-K_z + \Gamma_{xy} - \Gamma_{yz}) / \sqrt{6}, \notag\\
M_{ac} &= (2K_y - K_z + \Gamma_{xy} - 2\Gamma_{xz} + \Gamma_{yz}) / (3\sqrt{2}), \notag\\
M_{bc} &= (K_z + 2\Gamma_{xy} - 2\Gamma_{yz}) / (2\sqrt{3}).
\label{eq:couplings_global}
\end{align}

\begin{figure*}[!t]
\centering
\includegraphics[width=0.77\textwidth]{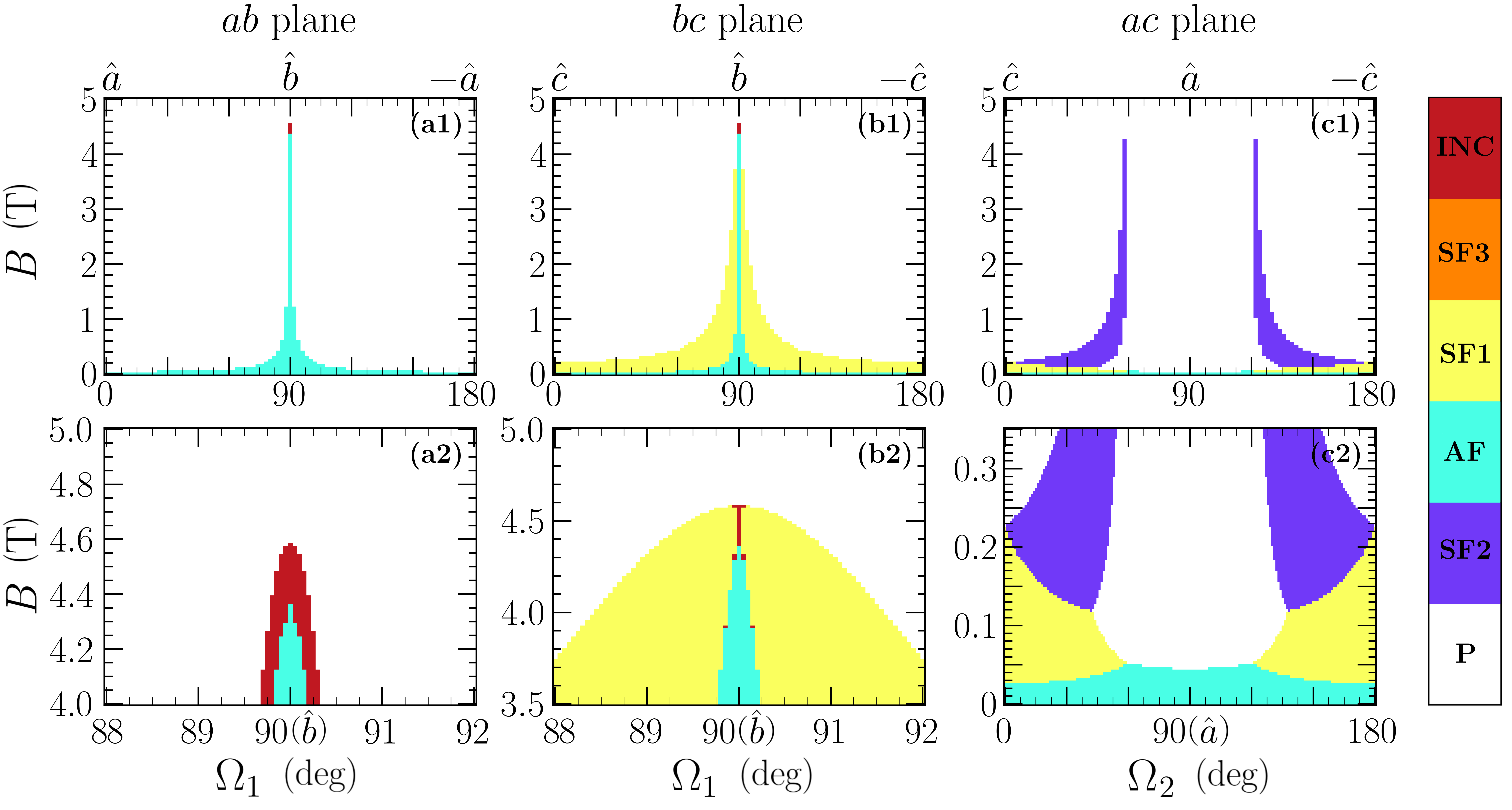}
\caption{
Model phase diagrams as function of field magnitude and direction, here for the parameter set of Woodland \textit{et al.} \cite{woodland23II}, to be compared to Fig.~\ref{fig:pd1}.
}
\label{fig:pdwood}
\end{figure*}

\section{Transverse-field Ising chain}

The one-dimensional Ising model in a transverse field,
\begin{equation}
\hat{H} = -J \sum_i (\sigma^z_i \sigma^z_{i+1} + \lambda \sigma^x_i)
\end{equation}
with Pauli matrices $\sigma^{x,z}_i$ representing spin-$1/2$ degrees of freedom, can be solved exactly, e.g., using a Jordan-Wigner transformation \cite{book_sachdev}. Its excitation spectrum can be computed to be $\epsilon_k = 2J \sqrt{1+\lambda^2-2\lambda\cos k}$. The spectrum is gapped for all $\lambda$ except at the quantum-critical point at $\lambda=1$.

In a semiclassical treatment, one can determine the dispersion of the spin-flip excitations of the field-polarized phase, $\lambda\gg 1$, using spin-wave theory. The result is $\epsilon_k = 2J \sqrt{\lambda^2-2\lambda\cos k}$, with the excitation gap closing at $\lambda=2$, i.e., twice the exact value. Hence, the semiclassical calculation overestimates the stability of the ordered phase against the transverse field. In the numerical results presented in this paper, we compensate for this by rescaling the transverse field components for each chain by a factor of two.
~\\[-12mm]
\section{Numerical results for the parameter set of Ref.~\onlinecite{woodland23II}}
\label{app:woodland}

While the body of the paper shows numerical results obtained for the set of model parameters from Ref.~\onlinecite{konieczna25}, we have also performed similar computations for the parameter set of Ref.~\onlinecite{woodland23II}. The resulting phase diagrams are shown in Fig.~\ref{fig:pdwood}. They largely agree with the ones in Fig.~\ref{fig:pd1}, with the following differences:

(i) The sequence of field-induced phases agrees for $\vec{B}\parallel\hat{b}$, but disagrees for $\vec{B}\parallel\hat{c}$ where the model of Ref.~\onlinecite{woodland23II} yields AF$\to$SF1$\to$P while that of Ref.~\onlinecite{konieczna25} predicts AF$\to$INC$\to$SF1$\to$SF2$\to$P, the latter in agreement with experiment. The sequences also disagree for $\vec{B}\parallel\hat{a}$, AF$\to$INC$\to$P vs. AF$\to$INC$\to$SF1$\to$P; but as noted above, the experiments \cite{weitzel00} reports SF3 instead of SF1.
(ii) Important quantitative differences concern the value of the critical field for $\vec{B}\parallel\hat{b}$ as well as the field width of the incommensurate phase. The latter is larger for the parameters of Ref.~\onlinecite{woodland23II} but still too small compared to experiment.
(iii) Similarly, the transition field into the high-field phase for $\vec{B}\parallel\hat{a}$ differs by a factor of three between the two computations; the results for the parameter set of Ref.~\onlinecite{konieczna25} appears closer to experiment \cite{weitzel00}.

The overall comparison with the experimental phase diagrams shows that the regions of INC and SF3 phases come out too small, reenforcing the notion that they are fluctuation-enhanced.



\begin{thebibliography}{99}
\bibitem{J_Hl_note}
The estimate $J_{\mathrm{H},\mathrm{l}} = 0.1$\,meV also appears roughly consistent with the fact that the zero-field N\'eel temperature is $T_{\mathrm{AF}}=1.97$\,K \cite{heid95} if one assumes ${J_{\mathrm{H},\mathrm{l}}\sim \frac{1}{2}k_{\mathrm{B}}T_{\mathrm{AF}}}$.
\bibitem{weitzel_note}
The model calculations of Ref.~\onlinecite{weitzel00} also predict that SF2 extends to large fields for intermediate directions in the $ac$ plane. Notably, their model is not based on pseudospins $1/2$, but includes the effect of higher CEF levels which effectively contain the alternating Ising axes.
\bibitem{koni_note}
Our notation in Appendix~\ref{app:coupling_matrices} slightly differs from that of Ref.~\onlinecite{konieczna25}: We use couplings
$(J_{\mathrm{H}}, K_y, K_z, \Gamma_{xy}, \Gamma_{xz}, \Gamma_{yz})$ instead of $(J_{\mathrm{K}}, \bar{K}, K, \Gamma_1, \Gamma_2, \Gamma_3)$.
\end{thebibliography}
\end{document}